\documentclass [a4paper, 11pt] {article}
\usepackage{amsmath}
\usepackage{amssymb}
\usepackage{amsfonts}
\usepackage{latexsym}
\usepackage{amsmath}
\usepackage{mathtools}
\usepackage{empheq}
\usepackage{graphicx}
\usepackage{caption}
\usepackage{subcaption}
\usepackage[toc,page]{appendix}
\usepackage{multicol}
\usepackage{color}
\begin{document}

\title{\textbf{Phase Mixing of Kink MHD Waves in the Solar Corona: Viscous Dissipation and Heating}}

\author{Zanyar Ebrahimi$^1$\thanks{E-mail: zebrahimi@maragheh.ac.ir},
Roberto Soler $^{2,3}$ and Kayoomars Karami $^4$\\
$^1$\tiny{Research Institute for Astronomy $\&$ Astrophysics of
Maragha, University of Maragheh, P. O. Box 55136-553, Maragheh, Iran}\\
$^{2}$\tiny{Departament de F\'{\i}sica, Universitat de les Illes Balears, E-07122, Palma de Mallorca, Spain}\\
$^3$\tiny{Institut d'Aplicacions Computacionals de Codi Comunitari ($IAC^3$), Universitat de les Illes Balears, E-07122, Palma de Mallorca, Spain}\\
$^4$\tiny{Department of Physics, University of Kurdistan, Pasdaran Street, P.O. Box 66177-15175, Sanandaj, Iran}}
\maketitle
\begin{abstract}
Magnetohydrodynamic (MHD) kink waves have been observed frequently in solar coronal flux tubes, which
makes them a great tool for seismology of the solar corona. Here, the effect of viscosity is studied on the evolution
of kink waves. To this aim, we solve the initial value problem for the incompressible linearized viscous MHD
equations in a radially inhomogeneous flux tube in the limit of long wavelengths. Using a modal expansion
technique the spatio-temporal behavior of the perturbations is obtained. We confirm that for large Reynolds
numbers representative of the coronal plasma the decrement in the amplitude of the kink oscillations is due to
the resonant absorption mechanism that converts the global transverse oscillation to rotational motions in the
inhomogeneous layer of the flux tube. We show that viscosity suppresses the rate of phase mixing of the
perturbations in the inhomogeneous region of the flux tube and prevents the continuous building up of small scales
in the system once a sufficiently small scale is reached. The viscous dissipation function is calculated to investigate
plasma heating by viscosity in the inhomogeneous layer of the flux tube. For Reynolds numbers of the order of
$10^6$–$10^8$, the energy of the kink wave is transformed into heat in two to eight periods of the kink oscillation. For
larger and more realistic Reynolds numbers, heating happens, predominantly, after the global kink oscillation is
damped, and no significant heating occurs during the observable transverse motion of the flux tube.
\end{abstract}


%________________________________________________________________________________________________________
\section{Introduction}
Aschwanden et al. (1999) and Nakariakov et al. (1999) were the first to identify the spatially resolved kink oscillations of coronal loops using the Transitional Region And Coronal Explorer (TRACE) observations in the 171 ${\AA}$ Fe IX emission lines. Because of the large number of observations of kink oscillations in coronal loops these oscillations are a great seismological tool to estimate the parameters of the solar coronal plasma such as the magnetic field, the plasma density, and the transport coefficients.

An interesting characteristic of kink oscillations in coronal loops is that they damp fast usually within 3-5 periods. Nakariakov et al. (1999) speculated that an anomalously high viscosity or resistivity in the coronal plasma could be responsible for this rapid damping. However, among the proposed mechanisms to justify the rapid damping of the kink waves (see e.g. Goossens et al. 2002; Ofman \& Aschwanden 2002; Ruderman \& Roberts 2002; Ofman 2005, 2009; Morton \& Erd\'{e}lyi 2009, 2010) resonant absorption is the strongest candidate, since, unlike other theories, resonant absorption is the only proposed mechanism that is able to explain short damping times (of the order of a few periods of the kink oscillation) without invoking anomalous processes or values of the dissipative coefficients many orders of magnitude larger than those expected in the corona. Resonant absorption is an ideal process that does not need strong diffusion to work. However, there is still no observational evidence for resonant absorption. This mechanism was first proposed by Ionson (1978) as a heating mechanism in coronal loops. Since then, many studies have developed the theory of resonant absorption (see e.g. Davila 1987; Sakurai et al. 1991a, 1991b; Goossens et al. 1995; Goossens \& Ruderman 1995; Erd\'{e}lyi 1997; Cally \& Andries 2010 among many others). The necessary condition in this mechanism is that the wave frequency lies in the local Alfv\'{e}n and/or slow frequency continuum.  In this situation the energy of the global mode oscillation transfers to the local perturbations in the inhomogeneous regions of the magnetic flux tube. As a result, the amplitude of the perturbations grows at the resonance point and the dissipation mechanisms become important in the resonance layer, where the oscillations make large gradients. Ruderman \& Roberts (2002) studied damping of kink oscillations in coronal loops. Considering the effect of viscosity in their analysis, they confirmed the previously obtained numerical result (see e.g. Poedts \& Kerner 1991; Tirry \& Goossens 1996) that the decay rate of the transverse oscillation is independent of the Reynolds number $R_v$ when $R_v\gg 1$. They concluded that Reynolds number affects only the perturbations in the resonance layer so that it is not possible to obtain the value of viscosity from the observations of decaying kink oscillations in coronal loops. Resonant absorption has been studied for various complex configurations of the magnetic flux tubes including curvature of the flux tube (Terradas et al. 2006), longitudinal density stratification (Andries et al. 2005; Karami \& Asvar 2007; Soler et al. 2011), twisted magnetic field (Karami \& Bahari 2010; Ebrahimi \& Karami 2016; Ebrahimi \& Bahari 2019) and magnetic field expansion (Shukhobodskiy et al. 2018; Howson et al. 2019). For a review on the theory of resonant absorption, see Goossens et al. (2011).

Another consequence of existing a continuum of Alfv\'{e}n frequencies in the flux tubes may be the phase mixing of the Alfv\'{e}n waves (Heyvaerts \& Priest 1983; Ireland \& Priest 1997; Karami \& Ebrahimi 2009; Prokopyszyn \& Hood 2019). In this mechanism due to inhomogeneity of the local Alfv\'{e}n phase speed across the background magnetic field the perturbations on different magnetic surfaces become out of phase as travel in the case of propagating wave or oscillate in the case of standing wave. In the developed stage of phase mixing even with a small amount of viscosity or resistivity the dissipation mechanisms become important and could transform the wave energy to heat. Ofman \& Aschwanden (2002) suggested that the loop oscillations are dissipated by phase mixing with viscosity of the order $\nu=10^{5.3\pm3.5} ~m^2s^{-1}$ that is anomalously many order of magnitudes higher than the classical coronal value of the shear viscosity, $\nu=1~ m^2s^{-1}$ (Ofman et al. 1994). It is believed that some small-scale turbulence and structure enhance the viscosity in the coronal loop plasma. Ofman et al. (1994) investigated the effect of viscous stress tensor on the heating of the corona by the resonant absorption and showed that the shear viscosity has the dominant role in the heating process but the compressive viscosity does not have a significant contribution (see also Erd\'{e}lyi \& Goossens 1995, 1996).

Goossens et al. (2014) investigated the nature of kink waves and stated that kink waves do not only involve purely transverse motions of solar magnetic flux tubes, but the velocity field is a spatially and temporally varying sum of both transverse and rotational motion. In an axisymmetric cylindrical flux tube, wave modes can be classified according to the value of the azimuthal wavenumber, $m$. In this paper we study modes with $m=1$, i.e., kink modes. The global kink mode is the only mode that is able to displace transversely (i.e., laterally) the axis of the cylinder. The global mode with $m=1$ is resonantly coupled to Alfv\'{e}n modes with $m=1$ in the nonuniform layer of the tube. These modes have both radial and azimuthal components of the displacement. Goossens et al. (2014) called the Alfv\'{e}n modes with $m\neq0$ as "rotational modes". The reason for calling these modes "rotational" is that their streamlines follow a closed curve, so that a fluid element flowing along one of those streamlines would describe a "rotation" around a certain point. In the case of $m=1$, the center of the rotation is not located on the axis of the cylinder (as happens for torsional motions with "m=0"), but at some place in between the axis and the boundary of the tube. Following Goossens et al. (2014), we use the term "transverse motion" to refer the lateral displacement of the flux tube caused by the global kink mode and the term "rotational motion" to refer the local Alfv\'{e}n perturbations inside the tube. Soler \& Terradas (2015, hereafter ST2015) investigated the resonant absorption of the kink MHD wave and phase mixing of its perturbations in coronal flux tubes. Using a modal expansion method they showed that the energy of the global kink oscillation of the flux tube is transformed into small-scale rotational motions in the nonuniform boundary of the tube which are eventually subject to the simultaneously occurring phase mixing process. However, ST2015 used the ideal MHD equations and did not consider the effect of dissipation terms in their analysis. At the developed stage of the phase mixing process where the perturbations are highly phase mixed, the dissipation mechanisms could suppress the rate of generating small scales in the system by coupling the perturbations on the neighboring magnetic surface and finally transform the kink wave energy to heat.

In this paper, our aim is to investigate the effect of viscosity on the kink MHD waves and show how viscosity modifies the previous results obtained by ST2015. Section \ref{model} presents the model and the governing MHD equations of motion. In section \ref{solution} we apply and extend the mathematical method used by ST2015 and give a solution to the equation of motion. The results are discussed in section \ref{results}. Finally we conclude the paper in section \ref{Conclusions}.

%________________________________________________________________________________________________________
\section{Equations of motion and model}\label{model}
We model a typical coronal loop by a straight cylinder that has a
circular cross section of radius $R$. The background
plasma density in cylindrical coordinates ($r$, $\varphi$, $z$) is
assumed to be as follows
\begin{equation}\label{rho}
\rho(r)=\left\{\begin{array}{lll}
    \rho_{{\rm i}},&r\leqslant r_1,\\
    \frac{\rho_{{\rm i}}}{2}\left[\left(1+\frac{\rho_{{\rm e}}}{\rho_{{\rm i}}}\right)-\left(1-\frac{\rho_{{\rm e}}}{\rho_{{\rm i}}}\right)\sin\left(\frac{\pi}{l}(r-R)\right)\right],&r_1< r< r_2,\\
    \rho_{{\rm e}},&r\geqslant r_2,
      \end{array}\right.
\end{equation}
where $r_1=R-l/2$ and $r_2=R+l/2$. Here, $l=r_2-r_1$ is the width of
the inhomogeneous region. The background magnetic field is assumed to be constant and aligned with the flux tube axis
everywhere, i.e. $\mathbf{B}=B_{{\rm 0}}\hat{z}$ where $B_0$ is constant.

The linearized MHD equations for an incompressible plasma with viscosity are as follows
\begin{equation}\label{mhd1}
     \rho(r)\frac{\partial^2 \boldsymbol{\xi}}{\partial t^2}=-\nabla p'+\frac{1}{\mu_0}(\nabla\times\mathbf {B'})\times{\mathbf B} +\rho(r)\nu\nabla^2\frac{\partial}{\partial t}\boldsymbol{\xi},
\end{equation}
\begin{equation}\label{mhd2}
    \mathbf{B'}=\nabla\times(\boldsymbol{\xi}\times\mathbf{B}),
\end{equation}
\begin{equation}\label{mhd3}
    \nabla\cdot\boldsymbol{\xi}=0,
\end{equation}
where $\boldsymbol{\xi}$ is the lagrangian displacement of the
plasma, $\mathbf{B'}$ and $p'$ are the Eulerian perturbations of the
magnetic field and plasma pressure, respectively. Here $\mu_0$ is the magnetic permeability
of the free space and $\nu$ is coefficient of the kinematic shear viscosity which is assumed to be uniform.
Using Eq. (\ref{mhd3}), we can rewrite Eq. (\ref{mhd2}) as
\begin{equation}\label{deltaB}
    \mathbf{B'}=(\mathbf{B}\cdot\nabla)\boldsymbol{\xi}.
\end{equation}
Substituting $\mathbf{B'}$ from Eq. (\ref{deltaB}) into (\ref{mhd1}) yields
\begin{equation}\label{mhd4}
     \rho(r)\frac{\partial^2 \boldsymbol{\xi}}{\partial t^2}=-\nabla P'+\frac{1}{\mu_0}B_0^2 \frac{\partial^2}{\partial z^2}\boldsymbol{\xi}+\rho(r)\nu\nabla^2\frac{\partial}{\partial t}\boldsymbol{\xi},
\end{equation}
or in a more compact form
\begin{equation}\label{mhd5}
      \mathcal{L}_A\boldsymbol{\xi}-\rho(r)\nu\nabla^2\frac{\partial}{\partial t}\boldsymbol{\xi}=-\nabla P',
\end{equation}
where $P'=p'+(\mathbf{B'}\cdot \mathbf{B})/\mu_0$ is the Eulerian perturbation of the total (gas plus magnetic) pressure and
\begin{equation}\label{LA}
      \mathcal{L}_A\equiv\rho(r)\frac{\partial^2}{\partial t^2}-\frac{1}{\mu_0}B_0^2 \frac{\partial^2}{\partial z^2},
\end{equation}
is the Alfv\'{e}n operator. In the absence of viscosity, Eq. (\ref{mhd5}) reduces to equation (8) of ST2015.
Since the equilibrium quantities are only functions of $r$, the perturbations can be Fourier-analyzed with respect to the $\varphi$ and $z$ coordinates. Hence,
\begin{equation}\label{pert}
    \begin{split}
       P'= & P'(r,t)~e^{i(m\varphi+k_z z)}, \\
       \boldsymbol{\xi}= & \boldsymbol{\xi}(r,t)~e^{i(m\varphi+k_z z)},
    \end{split}
\end{equation}
where $m$ is the azimuthal mode number and $k_z$ is the axial wave number. The three components of Eq. (\ref{mhd5}) and the incompressibility condition (Eq. \ref{mhd3}) form a system of 4 independent equations for $\xi_r$, $\xi_\varphi$, $\xi_z$ and $P'$ as follows
\begin{eqnarray}\label{system}
&&\mathcal{L}_A\xi_r-\rho(r)\nu\frac{\partial}{\partial t}\left(\nabla^2\xi_r-\frac{\xi_r}{r^2}-\frac{2}{r^2}\frac{\partial}{\partial\varphi}\xi_\varphi\right)= -\frac{\partial}{\partial r}P',\label{sys1}\\
&&\mathcal{L}_A\xi_\varphi-\rho(r)\nu\frac{\partial}{\partial t}\left(\nabla^2\xi_\varphi-\frac{\xi_\varphi}{r^2}+\frac{2}{r^2}\frac{\partial}{\partial\varphi}\xi_r\right)= -\frac{1}{r}\frac{\partial}{\partial \varphi}P',\label{sys2}\\
&&\mathcal{L}_A\xi_z-\rho(r)\nu\frac{\partial}{\partial t}\nabla^2\xi_z= -\frac{\partial}{\partial z}P',\label{sys3}\\
&&\frac{1}{r}\frac{\partial}{\partial r}\left(r\xi_r\right)+\frac{1}{r}\frac{\partial}{\partial \varphi}\xi_\varphi+\frac{\partial}{\partial z}\xi_z=0.\label{sys4}
\end{eqnarray}
Here, we apply $\frac{1}{r}\frac{\partial}{\partial \varphi}$ and $\frac{\partial}{\partial z}$ from left on Eqs. (\ref{sys2}) and (\ref{sys3}), respectively and add the resulting equations together. After that with the help of Eqs. (\ref{pert}) and (\ref{sys4}) we obtain $P'$ in terms of $\xi_r$ and $\xi_\varphi$ as
\begin{equation}\label{p}
    \begin{split}
        P'&=\frac{-1}{m^2/r^2+k_z^2}\left[\mathcal{L}_A\left(\frac{1}{r}\frac{\partial(r\xi_r)}{\partial r}\right)-\right.\\
        &\left.\rho(r)\nu\frac{\partial}{\partial t}\left(\nabla^2\left(\frac{1}{r}\frac{\partial}{\partial r}(r\xi_r)\right)+\left(\frac{2}{r^3}-\frac{2}{r^2}\frac{\partial}{\partial r}\right)\frac{\partial}{\partial \varphi}\xi_\varphi-\frac{2}{r^3}\frac{\partial^2}{\partial\varphi^2}\xi_r\right)\right].
    \end{split}
\end{equation}
Now we use the thin tube (TT) approximation in which the wavelength of the kink waves, $\lambda$, is much larger than the radius of the cross section of the flux tube, $R=(r_1+r_2)/2$, i.e. $Rk_z\ll 1$ or $R\ll\lambda$. The first two terms of Eq. (\ref{sys4}) have magnitudes of the order $\xi_0/R$ where $\xi_0$ is a typical value for the Lagrangian displacement. Since the characteristic length scale in the $z$ direction is $\lambda$, the order of the magnitude of the third term of Eq. (\ref{sys4}) is equal to $\xi_0/\lambda$. Hence, in TT approximation neglecting the third term of Eq. (\ref{sys4}) with respect to the first two terms, yields
\begin{equation}\label{xirxiphi}
  \frac{\partial}{\partial \varphi}\xi_\varphi\simeq-\frac{\partial}{\partial r}(r\xi_r).
\end{equation}
We use Eq. (\ref{xirxiphi}) to obtain a single equation for $\xi_r$. Since we use this approximation, it is not possible to obtain $\xi_z$ with respect to $\xi_r$ from Eqs. (\ref{sys1})-(\ref{sys4}). However, in the TT approximation, Goossens et al. (2009) showed that the longitudinal component of the displacement, $\xi_z$, is always much smaller than the other components and the dominant motion is in the horizontal plane normal to the background magnetic field. Substituting Eq. (\ref{xirxiphi}) in Eqs. (\ref{sys1}) and (\ref{p}) and eliminating $P'$ from the resulting equations, gives the equation for $\xi_r$ as follows
\begin{equation}\label{xir}
  \mathcal{L}_A\mathcal{L}_s\xi_r+\left(\frac{m^2}{r^2}+k_z^2\right)\frac{d\rho(r)}{dr}\frac{\partial^2}{\partial t^2}\frac{1}{r}\frac{\partial}{\partial r}(r\xi_r)=\nu\frac{\partial}{\partial t}\mathcal{L}_d\xi_r,
\end{equation}
where $\mathcal{L}_s$ is the surface wave operator (see ST2015 for more details) and $\mathcal{L}_d$ is the viscous damping operator which are defined as follows
\begin{equation}\label{Ls}
    \mathcal{L}_s\equiv\left(k_z^2+\frac{m^2}{r^2}\right)\frac{\partial^2}{\partial r^2}+\frac{1}{r}\left(k_z^2+\frac{3m^2}{r^2}\right)\frac{\partial}{\partial r}
    -\frac{1}{r^2}\left(k_z^2-\frac{m^2}{r^2}\right)-\left(k_z^2+\frac{m^2}{r^2}\right)^2,
\end{equation}
\begin{equation}\label{Ld}
\begin{split}
   \mathcal{L}_d\equiv & \left(\frac{m^2}{r^2}+k_z^2\right)^2\left[\rho(r)\left(\nabla^2-\frac{1}{r^2}+\frac{2}{r^2}\left(1+r\frac{\partial}{\partial r}\right)\right)\right. \\
     & \left.-\frac{\partial}{\partial r}\left(\frac{\rho(r)}{m^2/r^2+k_z^2}\left[\frac{\partial^3}{\partial r^3}+\frac{4}{r}\frac{\partial^2}{\partial r^2}-\left(\frac{m^2}{r^2}+k_z^2\right)\frac{\partial}{\partial r}-\frac{1}{r}\left(\frac{m^2+1}{r^2}+k_z^2\right)\right]\right)\right].
\end{split}
\end{equation}
In the absence of viscosity, Eq. (\ref{xir}) consistently reverts to Eq. (16) of ST2015. From Eq. (\ref{xirxiphi}) we find that $\xi_\varphi$ is related to $\xi_r$ as
\begin{equation}\label{xiphixir}
  i\xi_\varphi\simeq-\frac{1}{m}\frac{\partial}{\partial r}(r\xi_r).
\end{equation}
In this relation, the factor $i$ accounts for a phase difference of $\pi/2$ between $\xi_\varphi$ and $\xi_r$. So, for convenience, in order to avoid imaginary terms in the calculations, in the rest of the paper we redefine $i\xi_\varphi$ as $\xi_\varphi$.
%________________________________________________________________________________________________________
\section{Solution}\label{solution}
\subsection{Solution in the uniform regions ($r\leqslant r_1$, $r\geqslant r_2$)}
In the limit of small viscosity which is the case in the solar corona, we can neglect the effect of viscosity in the interior and exterior regions of the flux tube, because viscous effects are only important in the inhomogeneous regions where phase mixing operates (Heyvaerts \& Priest 1983). So, following ST2015 in TT approximation ($R k_z\ll 1$) solutions of $\xi_{r}$ representing the kink ($m=1$) waves in the constant density regions, i.e. $r\leqslant r_1$ and $r\geqslant r_2$,  are as follows
\begin{eqnarray}
    &&\xi_r(r,t)\approx A_{\rm i}(t),~~~~~~~r\leqslant r_1,\\
    &&\xi_r(r,t)\approx A_{\rm e}(t)r^{-2},~~r\geqslant r_2,
\end{eqnarray}
where $A_{\rm i}(t)$ and $A_{\rm e}(t)$ are the time-dependent amplitudes.

\subsection{Solution in the nonuniform region ($r_1<r<r_2$)}
 In the nonuniform region $r_1<r<r_2$, following ST2015, we perform a modal expansion of the radial component of the Lagrangian displacement $\xi_r(r,t)$ as
\begin{equation}\label{me}
    \xi_r(r,t)=\sum_{n=1}^{\infty}a_n(t)\psi_n(r).
\end{equation}
In cylindrical geometry, it is appropriate to set functions $\psi_n(r)$ as the orthogonal eigenfunctions of the
regular Sturm-Liouville system defined by the following Bessel differential equation
\begin{equation}\label{sturmliouville}
    \frac{{\rm d}^2\psi}{{\rm d}r^2}+\frac{1}{r}\frac{{\rm d}\psi}{{\rm d}r}+\left(\alpha^2-\frac{1}{r^2}\right)\psi=0.
\end{equation}
The boundary conditions are the continuity of the radial displacement of the plasma, $\xi_r$, and Lagrangian perturbation of total pressure, $\delta P=P'+\xi_r d P_0/d r$, at $r=r_1$ and $r=r_2$. Here, $P_0$ is the equilibrium total (gas + magnetic) pressure. For the equilibrium presented in this paper, $d P_0/d r=(1/\mu_0)(\mathbf{B}\cdot\nabla)\mathbf{B}=0$. Hence $\delta P=P'$ and the continuity of the Lagrangian perturbation of total pressure, $\delta P$, is satisfied by the continuity of the Eulerian perturbation of total pressure, $P'$. Neglecting the viscous term in Eq. (\ref{p}) at the boundaries of the inhomogeneous region i.e. $r=r_1$ and $r=r_2$, one can find that the remaining terms are proportional to $\xi_r$ or $\partial \xi_r/\partial r$. Hence, the continuity of $\delta P$ at the boundaries is satisfied with the continuity of $\xi_r$ and $\partial \xi_r/\partial r$. From the continuity of $\xi_r$ at $r=r_1$ and $r=r_2$ we obtain amplitudes of $\xi_r$ inside and outside the tube, i.e. $A_{\rm i}(t)$ and $A_{\rm e}(t)$, respectively as follows
\begin{equation}\label{b1}
    A_{\rm i}(t)=\left.\sum_{n=1}^{\infty}a_n(t)\psi_n(r)\right|_{r=r_1},
\end{equation}
\begin{equation}\label{b2}
    A_{\rm e}(t)=r^2\left.\sum_{n=1}^{\infty}a_n(t)\psi_n(r)\right|_{r=r_2}.
\end{equation}
From the continuity of $\partial \xi_r/\partial r$ at $r=r_1$ and $r=r_2$ we get
\begin{equation}\label{b3}
    \left.\sum_{n=1}^{\infty}a_n(t)\frac{d\psi_n(r)}{d r}\right|_{r=r_1}=0,
\end{equation}
\begin{equation}\label{b4}
    A_{\rm e}(t)=-\frac{r^3}{2}\left.\sum_{n=1}^{\infty}a_n(t)\frac{d\psi_n(r)}{d r}\right|_{r=r_2}.
\end{equation}
Subtracting Eq. (\ref{b4}) from Eq. (\ref{b2}) and multiplying the resulting equation by $2r^{-3}$ we get
\begin{equation}\label{b5}
    \left.\sum_{n=1}^{\infty}a_n(t)\left(\frac{2}{r}\psi_n(r)+\frac{d \psi_n(r)}{d r}\right)\right|_{r=r_2}=0.
\end{equation}
Since the functions $a_n(t)$ in Eqs. (\ref{b3}) and (\ref{b5}) are linearly independent, their coefficients must be zero, namely,
\begin{eqnarray}\label{bcpsi}
    \left.\frac{d \psi_n(r)}{d r}\right|_{r=r_1}=0,\\
   \left.\left(\frac{2}{r}\psi_n(r)+\frac{d \psi_n(r)}{d r}\right)\right|_{r=r_2}=0.
\end{eqnarray}
We use these equations as the boundary conditions governing $\psi_n(r)$ at $r=r_1$ and $r=r_2$.
Functions $\psi_n(r)$ also satisfy the following orthogonality condition
\begin{equation}\label{ort}
    \frac{1}{l}\int_{r_1}^{r_2}\psi_n(r)\psi_{n'}(r)r {\rm d}r=\delta_{nn'}~~~\forall\ n,n'\in\{1,2,3,\ldots\}.
\end{equation}
For more details on the solution of $\psi_n(r)$ see section 3.2 of ST2015.

In order to calculate the time dependent coefficients $a_n(t)$, we must truncate the
infinite series of Eq. (\ref{me}) to a finite number N of terms. Substituting Eq. (\ref{me}) into (\ref{xir}) we obtain
\begin{equation}\label{eqant}
\begin{split}
   &\left[\rho(r)\mathcal{L}_s\psi_n(r)+\left(\frac{m^2}{r^2}+k_z^2\right)\frac{d \rho}{d r}\left(\frac{\psi_n(r)}{r}+\frac{d \psi_n(r)}{d r}\right)\right]\frac{d^2 a_n(t)}{dt^2} \\
     & -\nu\mathcal{L}_D\psi_n(r)\frac{d a_n(t)}{dt}+\frac{B^2 k_z^2}{\mu}\mathcal{L}_s\psi_n(r)a_n(t)=0.
\end{split}
\end{equation}
Multiplying Eq. (\ref{eqant}) by $\psi_{n'}(r)$ and integrating the resulting equation over the interval $[r_1,r_2]$ we obtain the following matrix equation
\begin{equation}\label{eqant2}
  \mathbb{M}\ddot{\vec{a}}(t)+ \mathbb{G}\dot{\vec{a}}(t)+ \mathbb{H}\vec{a}(t)=0,
\end{equation}
where $\mathbb{M}$, $\mathbb{G}$ and $\mathbb{H}$ are square matrices of order $N$ defined as follows
\begin{eqnarray}
    &&M_{nn'}=\frac{1}{l}\int_{r_1}^{r_2}\left[\rho(r)\mathcal{L}_s\psi_{n'}(r)+\frac{{\rm d}\rho}{{\rm d}r}\left(k_z^2+\frac{m^2}{r^2}\right)
    \left(\frac{\psi_{n'}(r)}{r}+\frac{{\rm d}\psi_{n'}(r)}{{\rm d}r}\right)\right]\psi_n(r) r {\rm d}r,\\
    &&G_{nn'}=-\nu\frac{1}{l}\int_{r_1}^{r_2}\psi_n(r)\mathcal{L}_d\psi_{n'}(r)rdr,\\
    &&\begin{split}
    &H_{nn'}=k_z^2\frac{B_0^2}{\mu}\frac{1}{l}\int_{r_1}^{r_2}\psi_n(r)\mathcal{L}_s\psi_{n'}(r) r {\rm d}r,
    \end{split}
\end{eqnarray}
and $\vec{a}(t)$ is a column vector defined as
\begin{equation}\label{atvec}
  \vec{a}(t)=\left[a_1(t), a_2(t), \ldots, a_N(t)\right]^T,
\end{equation}
in which, the superscript $T$ denotes the transpose. The dot and double dot signs in Eq. (\ref{eqant2}) represent the first and the second derivative with respect to $t$, respectively. We rewrite Eq. (\ref{eqant2}) in the following form
\begin{equation}\label{gep1}
\left[
    \begin{matrix}
\emptyset & \mathbb{M} \\
\mathbb{M} & \mathbb{G}
\end{matrix}
\right]
\left[
\begin{matrix}
  \ddot{\vec{a}}(t) \\
  \dot{\vec{a}}(t)
\end{matrix}
\right]+
\left[
    \begin{matrix}
-\mathbb{M} & \emptyset \\
\emptyset & \mathbb{H}
\end{matrix}
\right]
\left[
\begin{matrix}
  \dot{\vec{a}}(t) \\
  \vec{a}(t)
\end{matrix}
\right]=0,
\end{equation}
where $\emptyset$ is the zero square matrix of order $N$. Using the following definitions
\begin{equation}\label{bnt1}
  \mathbb{A}\equiv\left[
    \begin{matrix}
\emptyset & \mathbb{M} \\
\mathbb{M} & \mathbb{G}
\end{matrix}
\right],~~
\mathbb{B}\equiv\left[
    \begin{matrix}
-\mathbb{M} & \emptyset \\
\emptyset & \mathbb{H}
\end{matrix}
\right],~~ \vec{b}(t)\equiv\binom{\dot{\vec{a}}(t)}{\vec{a}(t)},
\end{equation}
we can rewrite Eq.(\ref{gep1}) as
\begin{equation}\label{gep2}
  \mathbb{A} \dot{\vec{b}}(t)+ \mathbb{B} \vec{b}(t)=0,
\end{equation}
where $\mathbb{A}$ and $\mathbb{B}$ are the square matrices of order $2N$. By setting the temporal dependence of $b_n(t)$ as $\exp(\sigma t)$, Eq. (\ref{gep2}) can be cast in the form of a generalized eigenvalue problem, namely,
\begin{equation}\label{gep3}
  \sigma\mathbb{A}\vec{b}+ \mathbb{B} \vec{b}=0.
\end{equation}
By solving Eq. (\ref{gep3}) we obtain a set of $2N$ eigenvalues, $\sigma$, and the corresponding eigenvectors, $\vec{b}$. The time-dependent
coefficients, $b_n(t)$, can be expressed as a superposition of the eigenvectors as
\begin{equation}\label{bnt3}
  b_n(t)=\sum_{n'=1}^{2N}C_{n'}\beta_{nn'}e^{\sigma_{n'}t},
\end{equation}
where, $\beta_{nn'}$ is the $n$th component of the $n'$th eigenvector and $\sigma_{n'}$ is the $n'$th eigenvalue. The constant coefficients $C_{n'}$ are obtained from the initial conditions.
From the definition of $\vec{b}(t)$ in Eq. (\ref{bnt1}) we can see that the desired coefficients $a_n(t)$ correspond to the last $N$ components of $\vec{b}(t)$, i.e.
\begin{equation}\label{ant}
  a_n(t)=\sum_{n'=1}^{2N}C_{n'}\beta_{N+n,n'}e^{\sigma_{n'}t},~~~n=1, 2, \ldots, N.
\end{equation}
Hence the expression for $\xi_r(r,t)$ takes the following form
\begin{equation}\label{xir2}
    \xi_r(r,t)=\sum_{n=1}^{N}\sum_{n'=1}^{2N}C_{n'}\beta_{N+n,n'}e^{\sigma_{n'}t}\psi_n(r).
\end{equation}
\subsection{Initial Conditions}
As in ST2015 we take the initial conditions for $\xi_r$ as
\begin{eqnarray}
    &&\xi_r(r,t=0)=\left\{\begin{array}{lll}
    \xi_0,&r\leqslant r_1,&\\
    \xi_0\frac{\psi_1(r)}{\psi_1(r_1)},&r_1<r< r_2,&\\
    \xi_0\frac{\psi_1(r_2)}{\psi_1(r_1)}\left(\frac{r_2}{r}\right)^2,&r\geqslant r_2,
      \end{array}\right.\label{init1}\\
    &&\frac{\partial\xi_r}{\partial t}\Big|_{(r,t=0)}=0,\label{init2}
\end{eqnarray}
where $\xi_0$ is a constant. We choose these initial conditions in purpose in order to be sure that at time $t=0$ the entire energy of the perturbations is in the generalized Fourier mode with largest spatial scale i.e. $\psi_1(r)$. This enables us to investigate the process of phase mixing of the perturbations in which the energy of the wave transfers from large spatial scales to smaller ones. In other words, with these initial conditions, as time goes on, the Fourier modes with smaller and smaller spatial scales contribute in the evolution of the kink wave. So, the larger the number of the available modes, $N$, the larger the evolution time that we are allowed to proceed before the solutions become inaccurate (for more details see Cally 1991).

Here, we rewrite Eq. (\ref{bnt3}) in its matrix form, namely,
\begin{equation}\label{bnmat}
    \vec{b}(t)=\hat{\beta}e^{\Sigma t}\vec{C},
\end{equation}
where $\Sigma={\rm diag}(\sigma_1,\sigma_2,\ldots,\sigma_{2N})$, $\vec{C}=\left[C_1,C_2,\ldots,C_{2N}\right]^T$ and $\hat\beta$ is a $2N$ by $2N$ matrix that its columns are the eigenvectors of Eq. (\ref{gep3}). Evaluating Eq. (\ref{bnmat}) at $t=0$ yields
\begin{equation}\label{bnmatt0}
    \vec{b}(t=0)=\hat{\beta}\hat{I}\vec{C}=\hat{\beta}\vec{C},
\end{equation}
where $\hat{I}$ is the identity matrix of size $2N\times 2N$. Assuming that the matrix $\hat{\beta}$ has a non-zero determinant, we obtain the coefficient vector $\vec{C}$ as follows
\begin{equation}\label{vecc}
    \vec{C}=\hat{\beta}^{-1}\vec{b}(t=0)=\hat{\beta}^{-1}\left[\begin{matrix}
  \dot{\vec{a}}(t=0) \\
  \vec{a}(t=0)
\end{matrix}\right].
\end{equation}
Setting $t=0$ in Eq. (\ref{me}) and using the initial conditions presented in Eqs. (\ref{init1}) and (\ref{init2}) one can easily find that the only non-zero component of $\vec{b}(t=0)$ is $b_{N+1}(t=0)=a_1(t=0)=\xi_0/\psi_1(r_1)$. Hence the coefficients $C_n$ are obtained as
\begin{equation}\label{cn}
    C_n=\beta^{-1}_{n,N+1}\frac{\xi_0}{\psi_1(r_1)}.
\end{equation}

%________________________________________________________________________________________________________
\section{Numerical results }\label{results}

In order to solve Eq. (\ref{gep3}), the density ratio of the flux tube is considered to be $\rho_{\rm i}/\rho_{\rm e}=5$. The observational values of this parameter has been reported to be in the range [2, 10] (Aschwanden et al. 2003). The azimuthal mode number representing the kink waves is $m=\pm1$. For the model considered in this paper, the results for both $m=+1$ and $m=-1$ are the same. Here, we take $m=+1$. Since we use the TT approximation, the longitudinal wavenumber is assumed to be $k_z=\frac{\pi}{100R}$. For the thickness of the inhomogeneous layer we consider two cases $l/R=0.2$ and $l/R=1$ that represent a thin and a thick layer, respectively. To consider the effect of viscosity, it is appropriate to use a dimensionless quantity, the viscous Reynolds number which is defined as
\begin{equation}\label{Reynolds}
    R_v\equiv\frac{lv_{\rm Ai}}{\nu}.
\end{equation}
Here, the Alfv\'{e}n speed $v_{\rm Ai}=B_0/\sqrt{\mu_0\rho_{\rm i}}$ is the characteristic speed for the propagation of kink waves in the flux tube.
Traditional value of the coefficient of shear viscosity in the solar corona is of the order $\nu=1~m^2/s$ (see Ofman et al. 1994 and references therein). With $l=10^6 m$ and $v_{\rm Ai}=10^6 m/s$, the corresponding Reynolds number is $R_v=10^{12}$. Due to computational limitations, we are forced to consider smaller $R_v$ in our results. However, the conclusions we obtain can be easily generalized to the case of larger $R_v$. When appropriate, we shall stress what differences would appear if more realistic values of $R_v$ were considered.
% \textit{As stated by Ofman et al. (1994), in the solar corona, heating due to the compressive viscosity is negligible compared to the shear viscosity dissipation. It has been shown by Mok (1987) that under certain
%solar conditions, such as in active regions, terms proportional to $\eta_1$ (using Braginskii’s [1965] notation) in the stress tensor are
%dominant, although $\eta_0$ is numerically many orders of magnitude larger.}
%Following Braginskii (1965) (see also Ofman et al. 1994) the coefficient of shear viscosity of a magnetized plasma is obtained from the following relation
%\begin{equation}\label{spitzer}
%    \nu= \frac{3nk_B T}{10\rho\omega_c^2\tau}~m^2 s^{-1},
%\end{equation}
%where $T$ is the temperature in Kelvin, $\rho$ is the plasma density in $kg.m^{-3}$. The bright loops in the active region have  typical densities of the order $10^{15}$ to $10^{16} m^{-3}$. For a plasma with temperature $T=10^6 K$ and density $n=10^{15} m^{-3}$ . Putting these parameters in Eq. (\ref{spitzer}) gives the viscosity as $\nu=6.8\times 10^{13} cm^2 s^{-1}$. Assuming $B_0=10 G$ for the magnetic field inside a typical coronal magnetic flux tube, yields $v_{\rm Ai}\simeq 690 km s^{-1}$. Hence for a thin transitional layer with $l/R=0.2$ the viscous Reynolds number is $R_v\simeq 20$.

\begin{figure}
  \centering
  \begin{tabular}{ccc}
    %\includegraphics[width=70mm]{spec5_02.eps}& \includegraphics[width=70mm]{spec5_1.eps}\\
    \includegraphics[width=70mm]{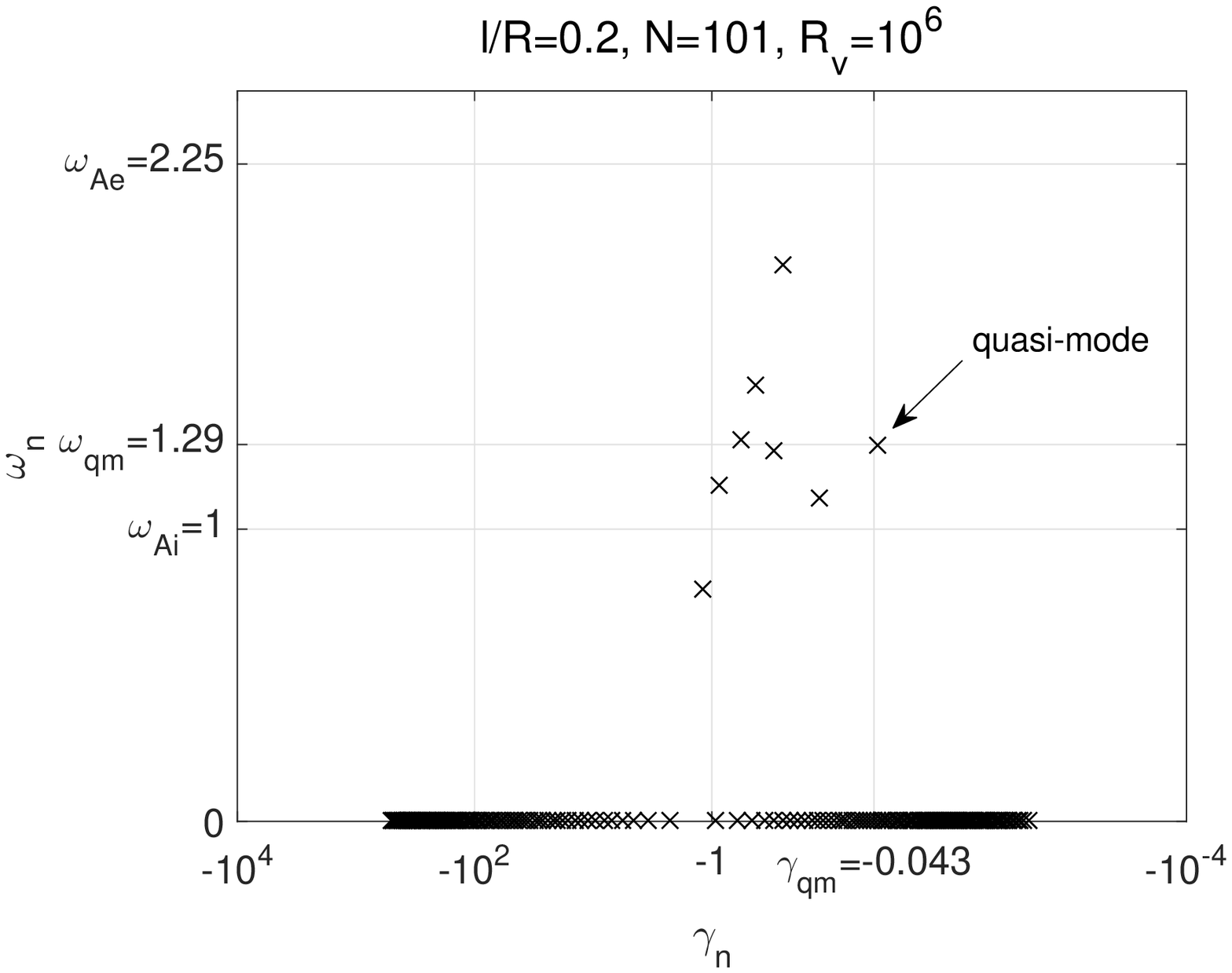}& \includegraphics[width=70mm]{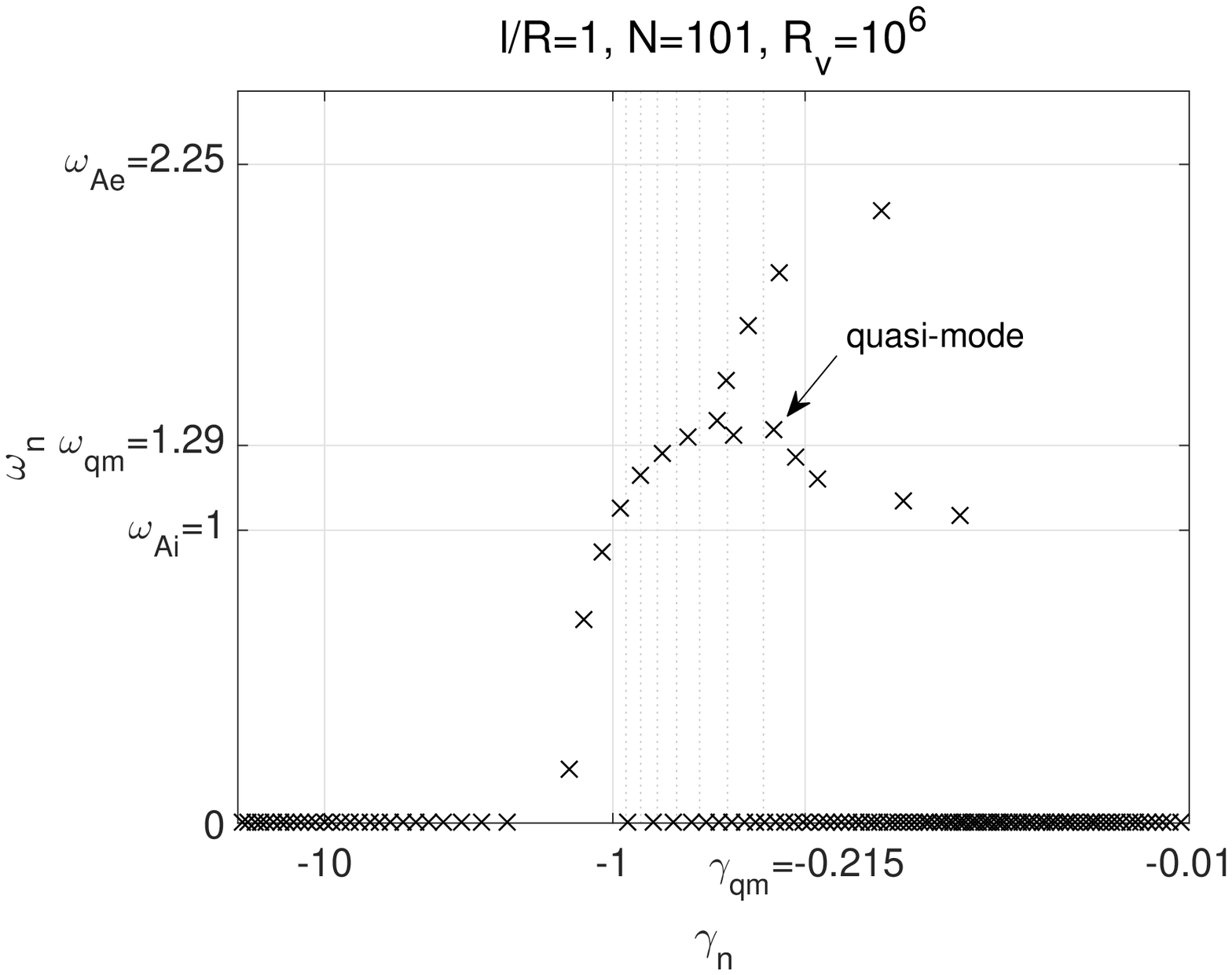}\\
    \includegraphics[width=70mm]{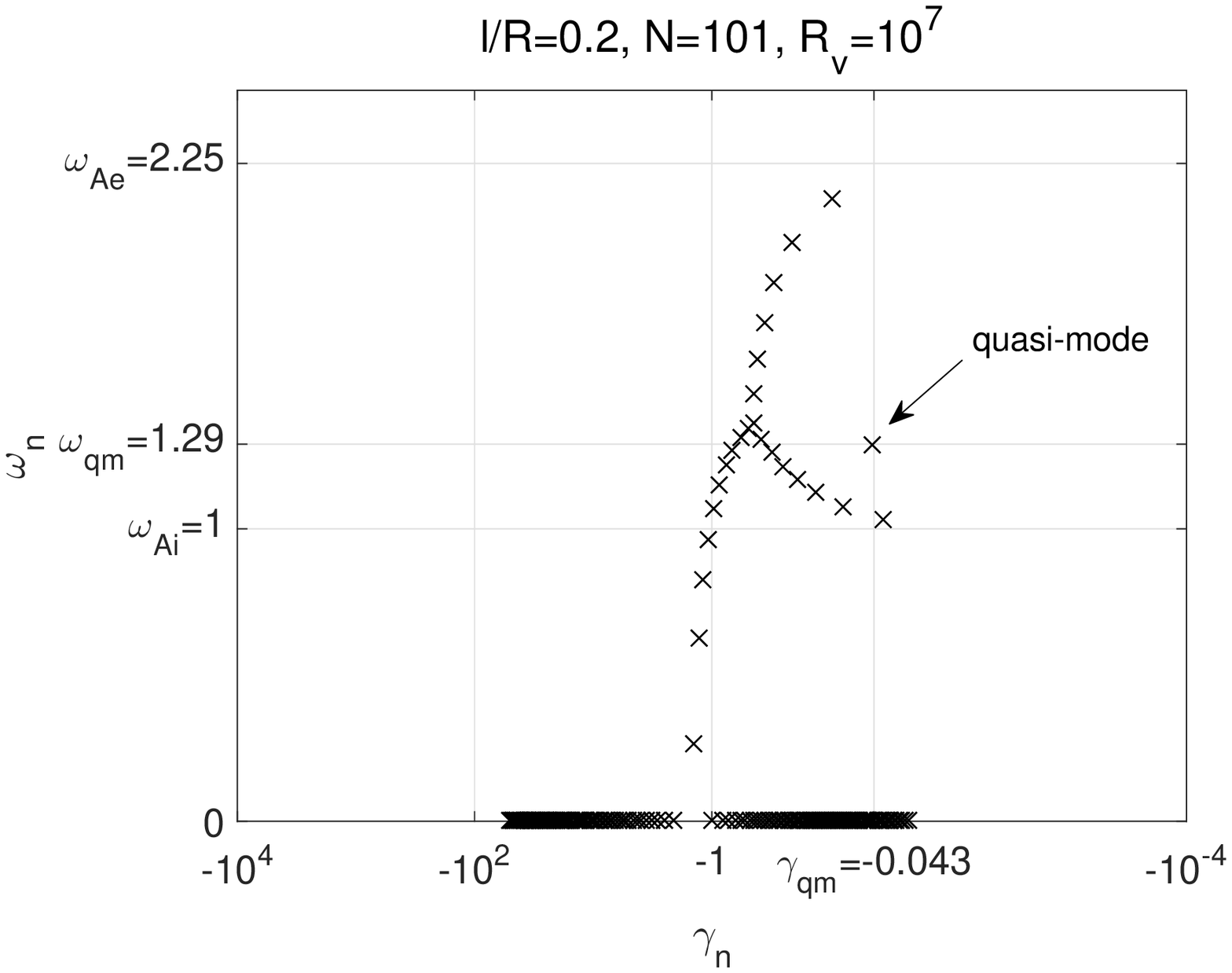}& \includegraphics[width=70mm]{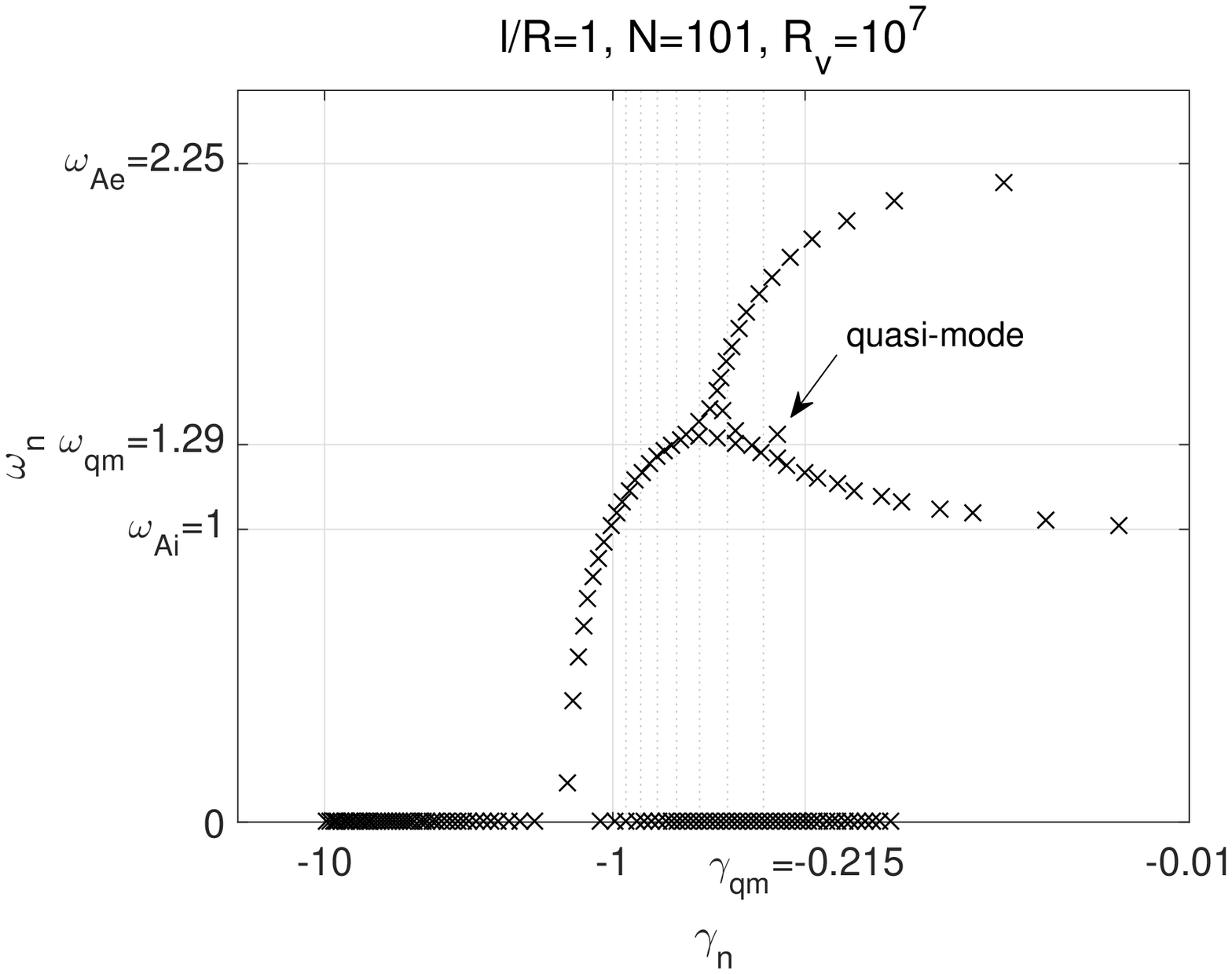}
  \end{tabular}
\caption {Spectrum of the eigenvalues for $k_z=\pi/100$ and $\rho_{\rm i}/\rho_{\rm e}=5$. Left and right panels are for $l/R=0.2$ and $l/R=1$, respectively. Top and bottom panels are for $R_v=10^6$ and $R_v=10^7$, respectively.}
    \label{spec}
\end{figure}

Solving the generalized eigenvalue problem (\ref{gep3}) results to 2N eigenvalues, $\sigma\equiv-i\tilde\omega$ where $\tilde\omega\equiv\omega+i\gamma$ is the complex eigenfrequency. These eigenvalues are real or come in pairs $(\sigma,\sigma^*)$ where $\sigma^*$ is the complex conjugate of $\sigma$. Figure \ref{spec} shows the $\omega_n\geqslant 0$ part of the spectrum of the complex eigenvalues $\sigma_n\equiv -i\omega_n+\gamma_n$ with $N=101$ for $l/R=0.2$ and $l/R=1$ and Reynolds numbers $R_v= 10^6, 10^7$. Here, $\omega_n$ and $\gamma_n$ are the frequency and the damping rate of the $n$'th eigenmode, respectively. In the figure, $\omega_{\rm Ai}=k_z B_0/\sqrt{\mu_0\rho_{\rm i}}$ and $\omega_{\rm Ae}=k_z B_0/\sqrt{\mu_0\rho_{\rm e}}$ are the Alfv\'{e}n frequencies inside and outside of the tube, respectively. Note that the frequencies and damping rates are in units of the internal Alfv\'{e}n frequency, $\omega_{\rm Ai}$. The complex spectrum has the typical three-branch structure found in previous papers that computed the resistive Alfv\'{e}n spectrum in similar configurations (see Poedts \& Kerner 1991). It is clear from Figure \ref{spec} that the smaller the value of the Reynolds number (larger coefficient of viscosity) the more number of modes with $\omega=0$ are present in the complex spectrum. This result is in well agreement with  previous results obtained in resistive MHD (see e.g. Van Doorsselaere and Poedts 2007).  An interesting result is that similar to the resistive MHD analysis (see e.g. Poedts and Kerner 1991) in viscous MHD, one of the eigenvalues of the complex spectrum could be identified as the damped quasi-mode solution (global mode) of ideal MHD. The real part of the frequency of this solution is the kink mode frequency, and the imaginary part is its corresponding damping rate due to the resonant absorption mechanism. Following Ruderman and Roberts (2002) and Goossens et al. (2002), the quasi-mode frequency $\omega_{\rm qm}$ and damping rate $\gamma_{\rm qm}$ for MHD kink waves in the TT and thin boundary (TB) ($l/R\ll1$) approximations are
\begin{equation}\label{omegaqm}
    \omega_{\rm qm}\simeq\omega_{\rm k}=\sqrt{\frac{2\rho_{\rm i}}{\rho_{\rm i}+\rho_{\rm e}}}\omega_{\rm Ai},
\end{equation}
\begin{equation}\label{gammaqm}
    \gamma_{\rm qm}=-\omega_{\rm k}\frac{l}{4R}\frac{\rho_{\rm i}-\rho_{\rm e}}{\rho_{\rm i}+\rho_{\rm e}}.
\end{equation}
From these equations for $\rho_{\rm i}/\rho_{\rm e}=5$ the quasi-mode frequency is obtained as $\omega_{qm}=1.29\omega_{\rm Ai}$. Also the quasi-mode damping rates for $l/R=0.2$ and $l/R=1$ are $\gamma_{\rm qm}=-0.043\omega_{\rm Ai}$ and $\gamma_{\rm qm}=-0.215\omega_{\rm Ai}$, respectively. As illustrated in Figure \ref{spec}, the singled out mode matches the quasi-mode solution especially for the thin transitional layer case, $l/R=0.2$, and larger values of the Reynolds number. For instance, for $l/R=0.2$ and $R_v=10^7$ the singled out eigenfrequency is $\tilde\omega=1.29-0.044i$ that is in well agreement with the result obtained from the above analytic approximations. For $l/R=1$ and $R_v=10^7$ the corresponding eigenfrequency is $\tilde\omega=1.33-0.269i$. Hence, in the case of thick transitional layer, the numerical results and the analytic approximations deviate more from each other. This is consistent with the fact that the analytic approximations are only strictly valid when $l/R \ll 1$. For a study of the validity of the approximations beyond its theoretical range of applicability see Soler et al. (2014). These results show that for the kink waves, the quasi-mode solution of the ideal MHD can be identified as an eigenmode of the viscous MHD spectrum. This correspondence is very clear in the case of thin nonuniform layers and not very large Reynolds numbers. However, as the thickness of the layer or the Reynolds numbers increase, the identification of the quasi-mode in the dissipative spectrum is more confusing since the quasi-mode gets embedded in one of the branches of the spectrum and becomes indistinguishable from an ordinary dissipative Alfv\'{e}n mode (see discussions on this issue in Van Doorsselaere \& Poedts 2007 and  Soler et al. 2013). A detailed analysis of the peculiar behaviour of the quasi-mode in the complex spectrum is beyond the purpose of the present paper.

%\begin{figure}
%   \centering
%    \includegraphics[width=70mm]{qm_versus_Rv.eps}
%\caption {}
%    \label{qm_versus_Rv}
%\end{figure}

Once the dissipative spectrum is computed, the time-dependent behavior of the perturbations is obtained by the superposition of all the modes in the spectrum according to the prescribed initial condition, which represents a transverse, i.e., kink displacement of the whole tube (see Section 3.3). In short, the evolution of the subsequent global kink oscillation is determined by two simultaneously working mechanisms: resonant absorption and phase mixing. On the one hand, resonant absorption is responsible for a radial flux of energy towards the nonuniform layer of the flux tube, and its net effect is producing the damping of the global kink oscillation. As a result, the amplitude of the displacement at the tube axis decreases in time. On the other hand, the energy accumulated at the nonuniform layer because of resonant absorption drives local Alfv\'{e}n waves with the same azimuthal symmetry as the original kink oscillation, i.e., $m=1$. Although these Alfv\'{e}n waves have both radial and azimuthal components of the displacement, they are largely polarized in the azimuthal direction. The Alfv\'{e}n waves undergo the process of phase mixing because of the spatially-dependent Alfv\'{e}n velocity. This causes the building up of small scales in the nonuniform layer and the subsequent energy cascade to these small scales. Then, viscous dissipation becomes important. In the following paragraphs, we analyze the dynamics we have just summarized.

\begin{figure}
  \centering
  \begin{tabular}{ccc}
    \includegraphics[width=50mm]{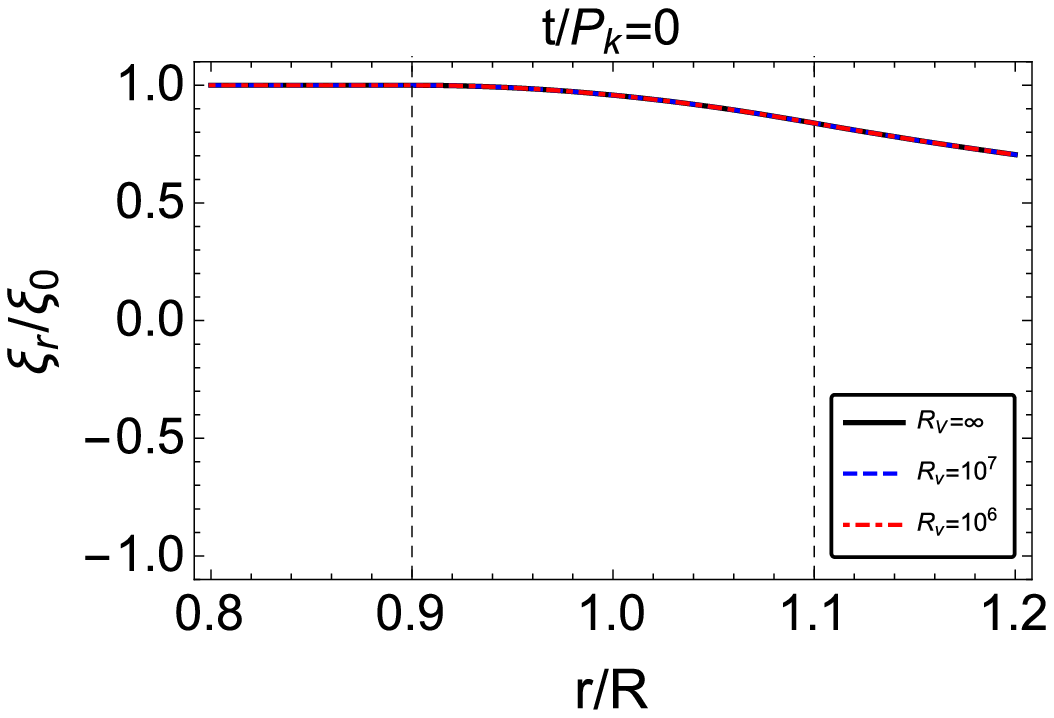}& \includegraphics[width=50mm]{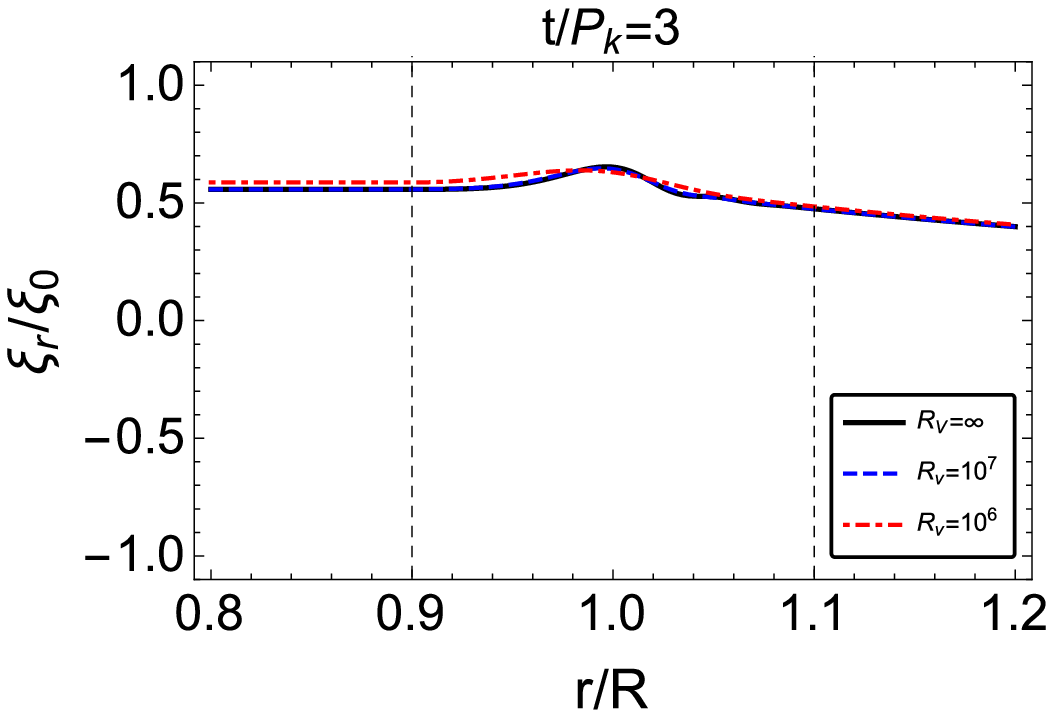}& \includegraphics[width=50mm]{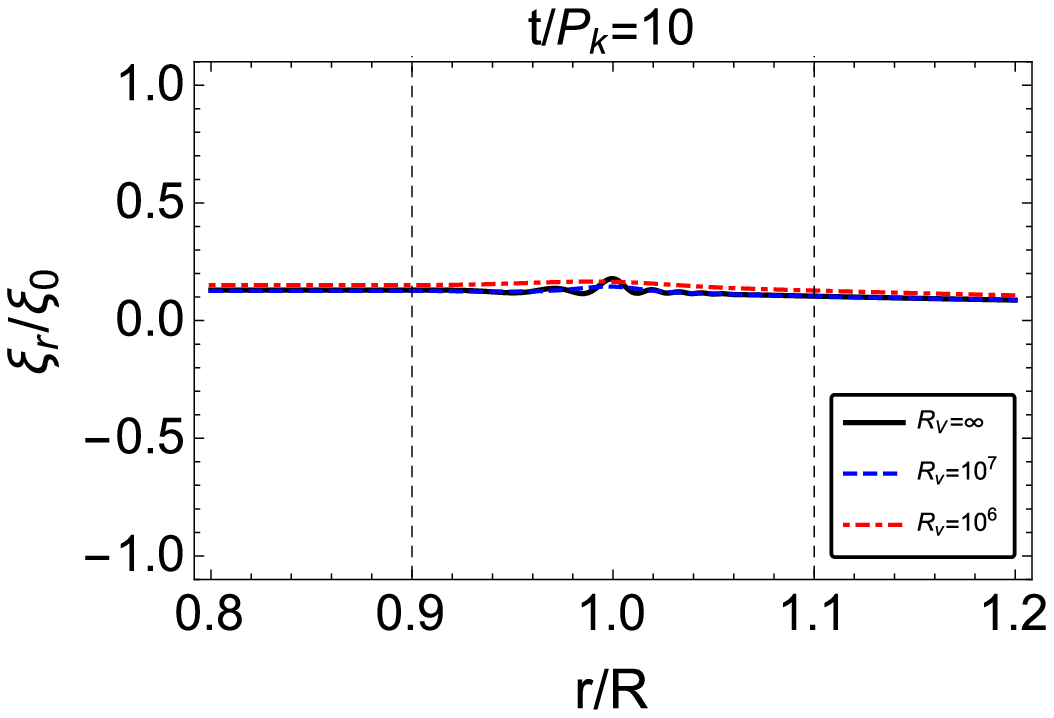}\\
     \includegraphics[width=50mm]{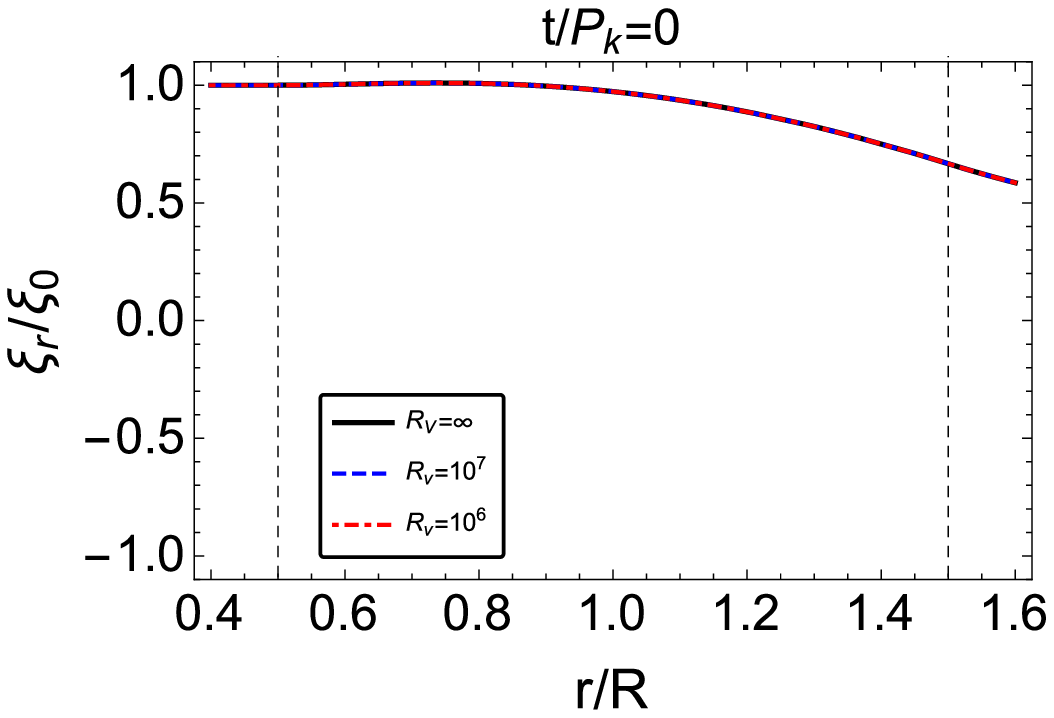}& \includegraphics[width=50mm]{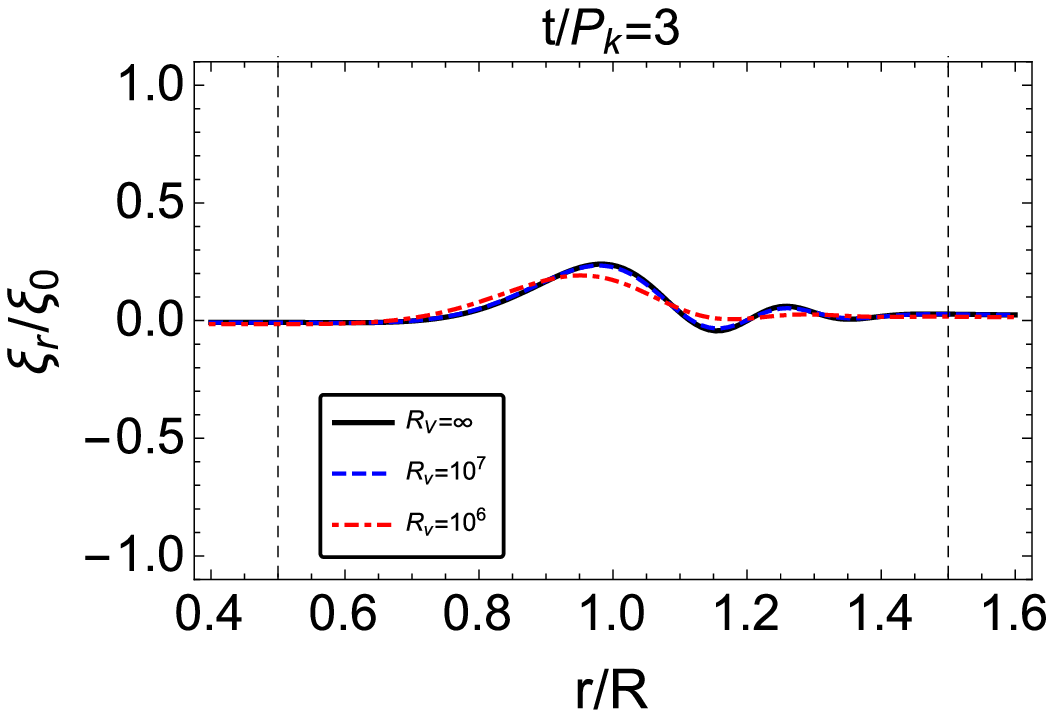}&  \includegraphics[width=50mm]{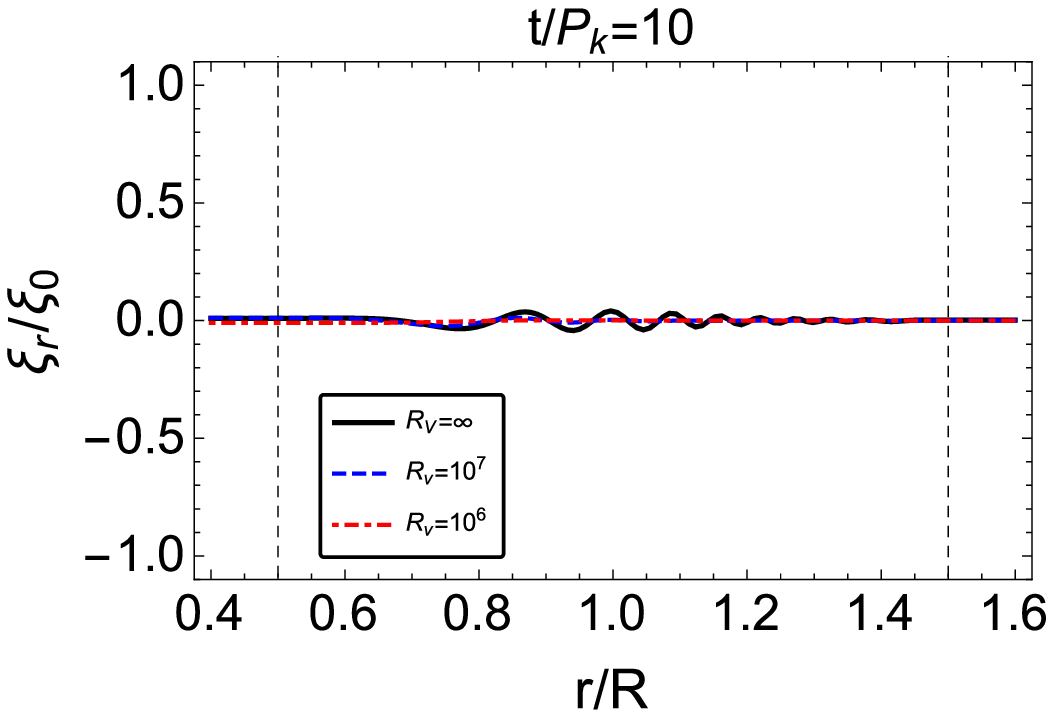}
  \end{tabular}
\caption {Evolution of $\xi_r$ in a nonuniform tube with $l/R=0.2$ (top panels) and $l/R=1$ (bottom panels) for $R_v=\infty, 10^6$ and $10^7$. Left, center and right panels denote $t/P_k = 0$, $3$ and $10$, respectively. The left and right vertical dashed lines locate $r=r_1$ and $r=r_2$, respectively. Other auxiliary parameters are as in Figure \ref{spec}. An animation of this Figure is available.}
    \label{xirr}
\end{figure}

\begin{figure}
  \centering
  \begin{tabular}{ccc}
    \includegraphics[width=50mm]{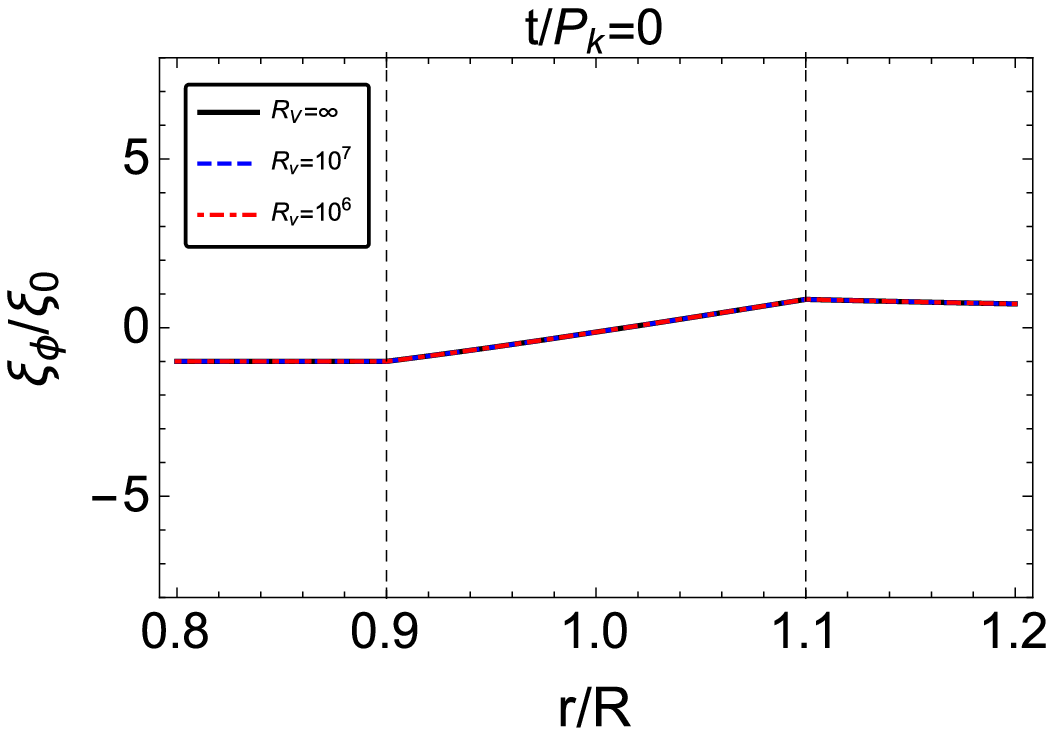}& \includegraphics[width=50mm]{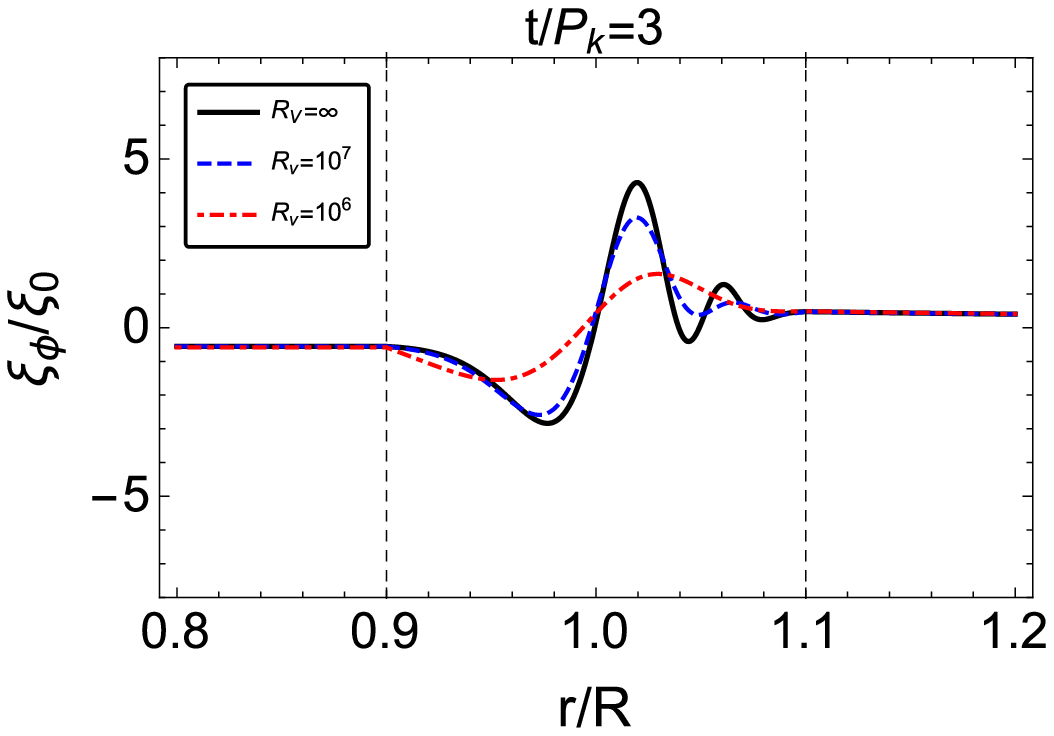}& \includegraphics[width=50mm]{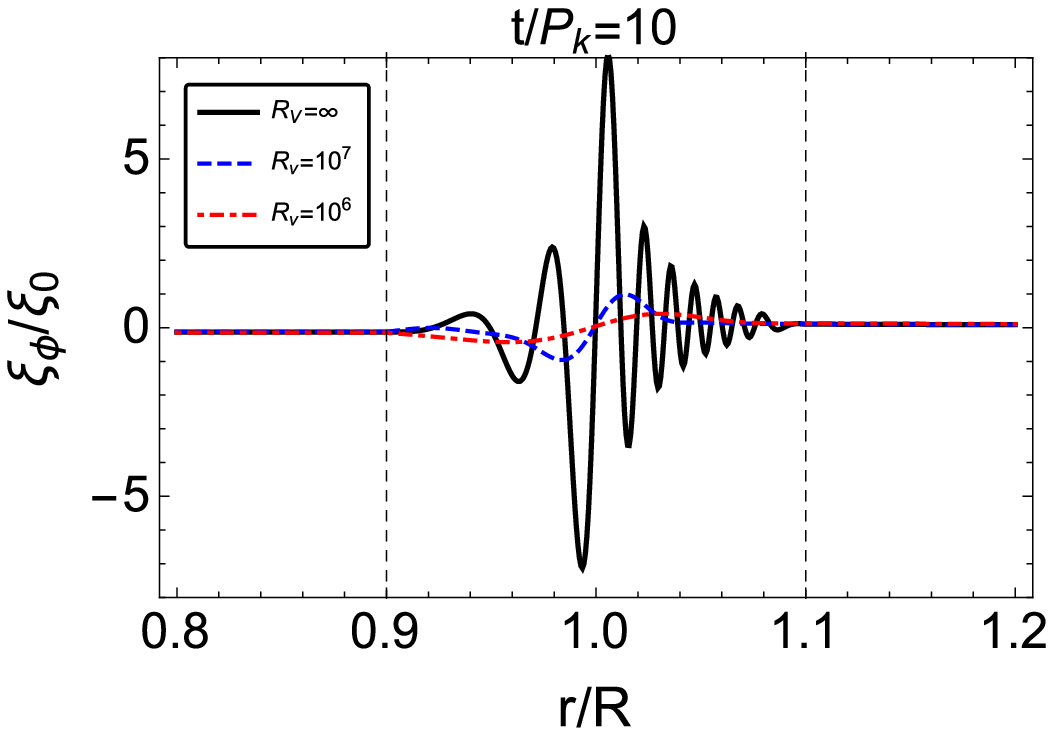}\\
    \includegraphics[width=50mm]{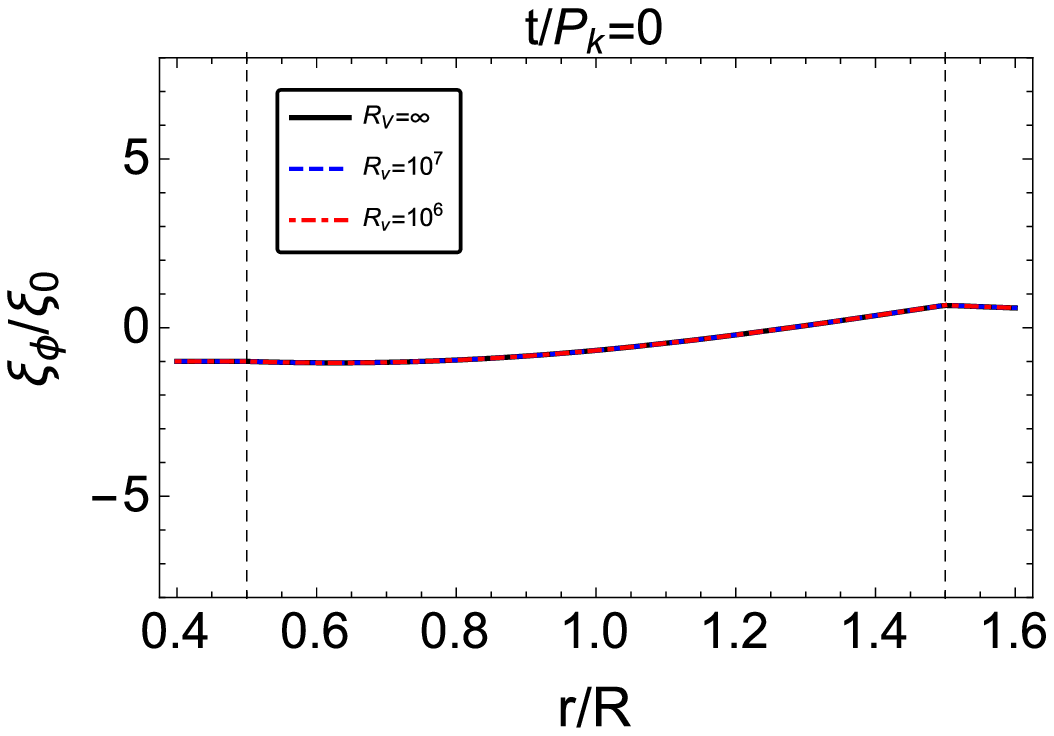}&  \includegraphics[width=50mm]{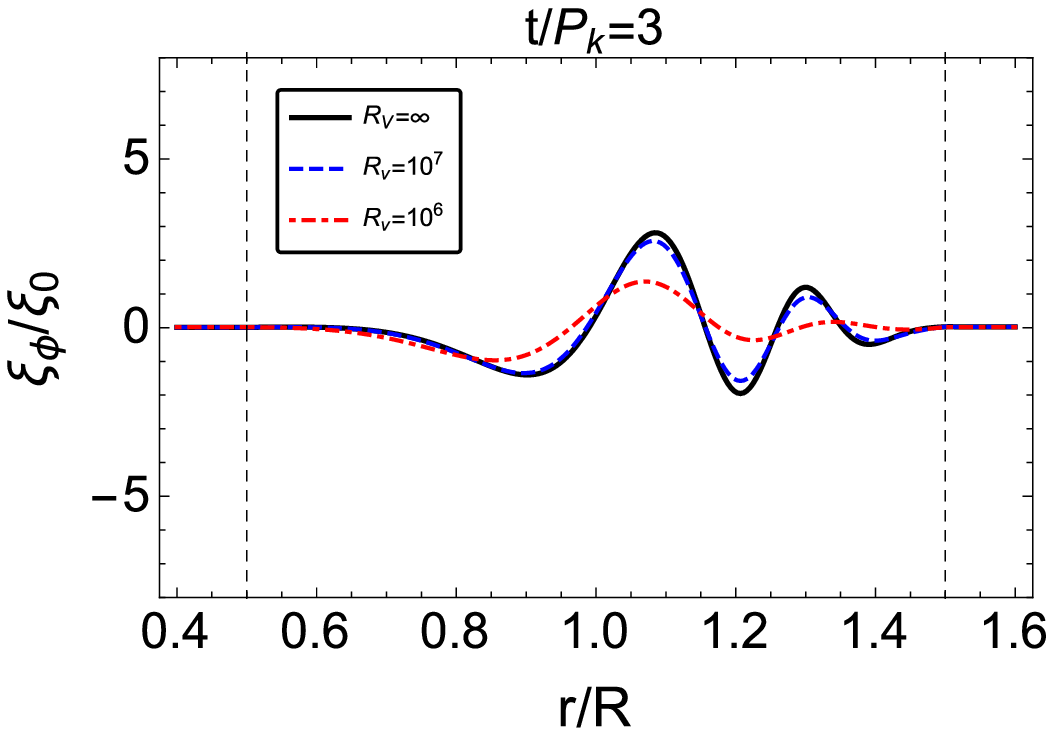}& \includegraphics[width=50mm]{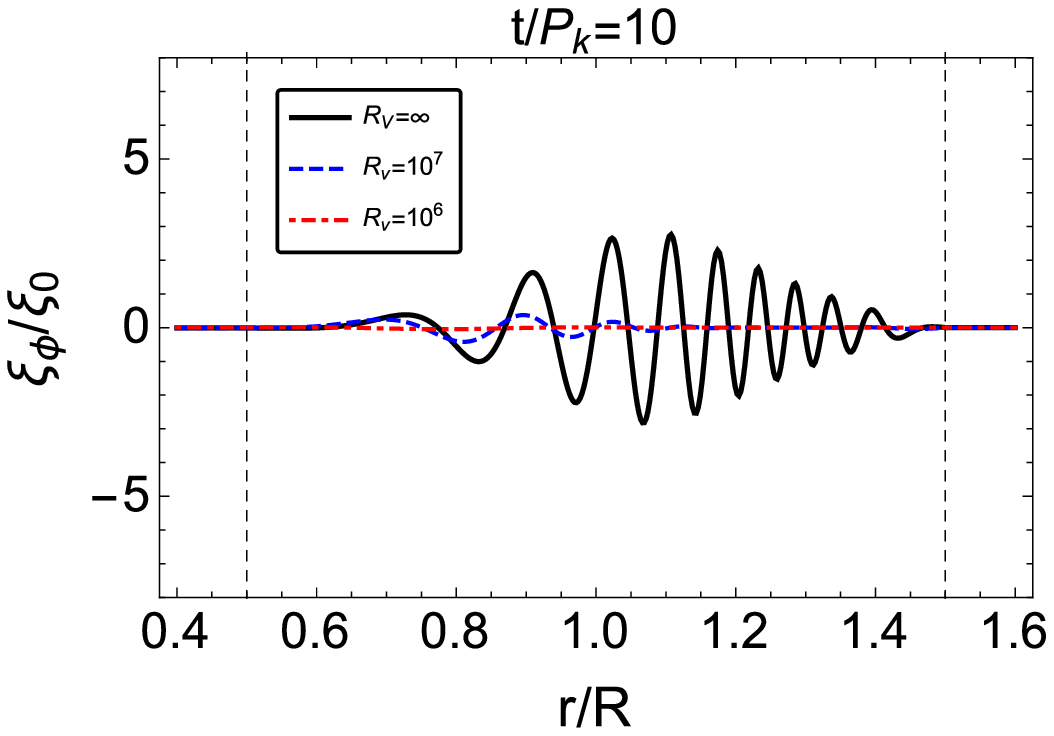} \\
  \end{tabular}
\caption {Same as Figure \ref{xirr} but for $\xi_\varphi$. An animation of this Figure is available.}
    \label{xiphir}
\end{figure}

Figures \ref{xirr} and \ref{xiphir} show the evolution of $\xi_r$ and $\xi_\varphi$, respectively. Time is in units of the period of the kink oscillation in TTTB approximations, $P_{\rm k}=2\pi/\omega_{\rm k}$. In Figures \ref{xirr} and \ref{xiphir} the solid black curves represent the results previously obtained by ST2015 in the absence of viscosity, i.e. $R_v=\infty$. The blue dashed and red dashed-dotted curves are for $R_v=10^7$ and $R_v=10^6$, respectively. Figures reveal that in the presence of viscosity, unlike the results obtained in ideal MHD (see Soler and Terradas 2015; Ebrahimi et al. 2017) perturbations are not allowed to be phase mixed indefinitely. The smaller the Reynolds number, the quicker the dissipative solution departs from the ideal solution. The existence of viscosity cause to coupling of the perturbations on the neighboring magnetic surfaces. This effect suppresses the phase mixing when a certain spatial scale is reached and transforms the total (kinetic plus magnetic) energy of the perturbations to heat via dissipation.

\begin{figure}
  \centering
  \begin{tabular}{ccc}
    \includegraphics[width=50mm]{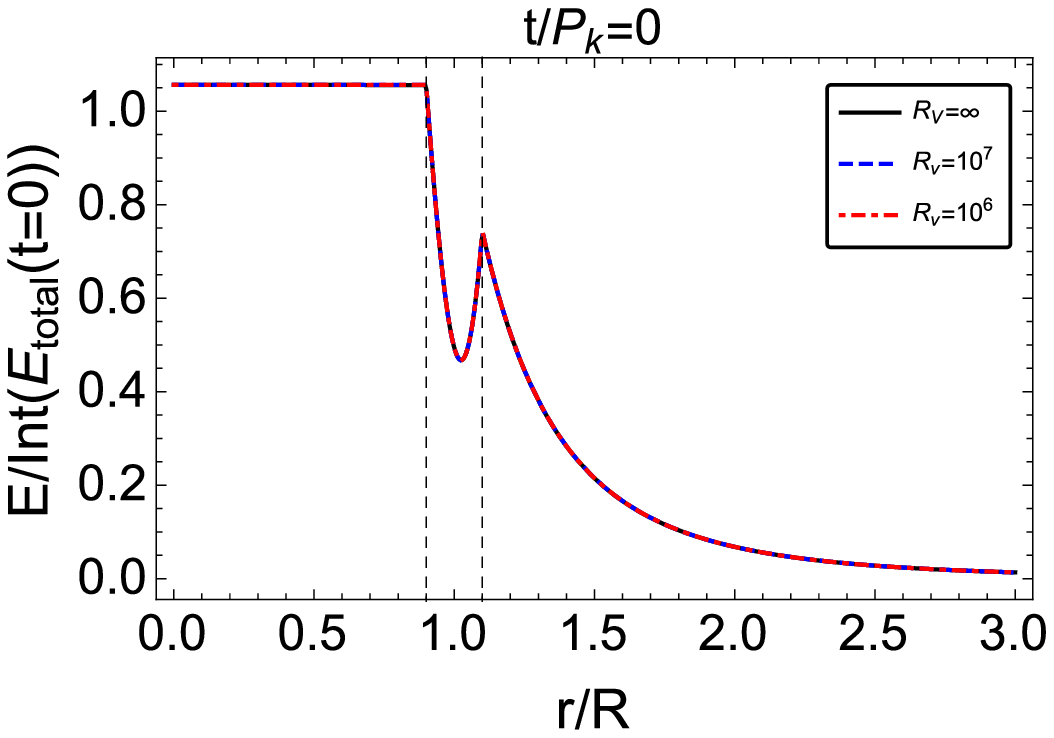}& \includegraphics[width=50mm]{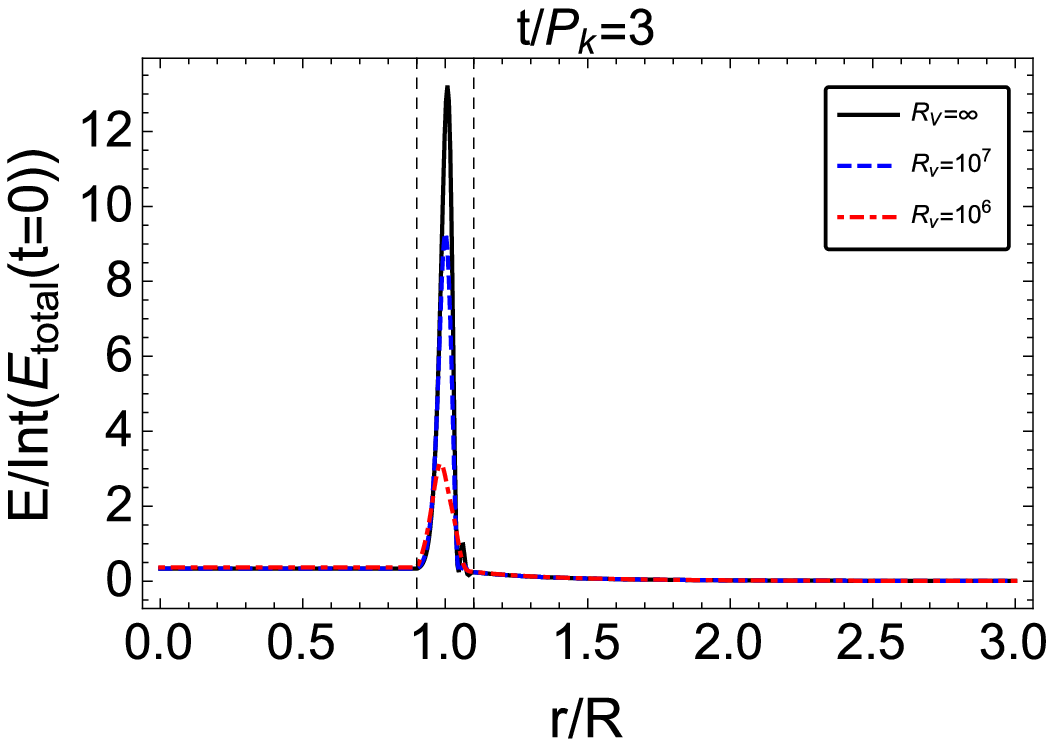}& \includegraphics[width=50mm]{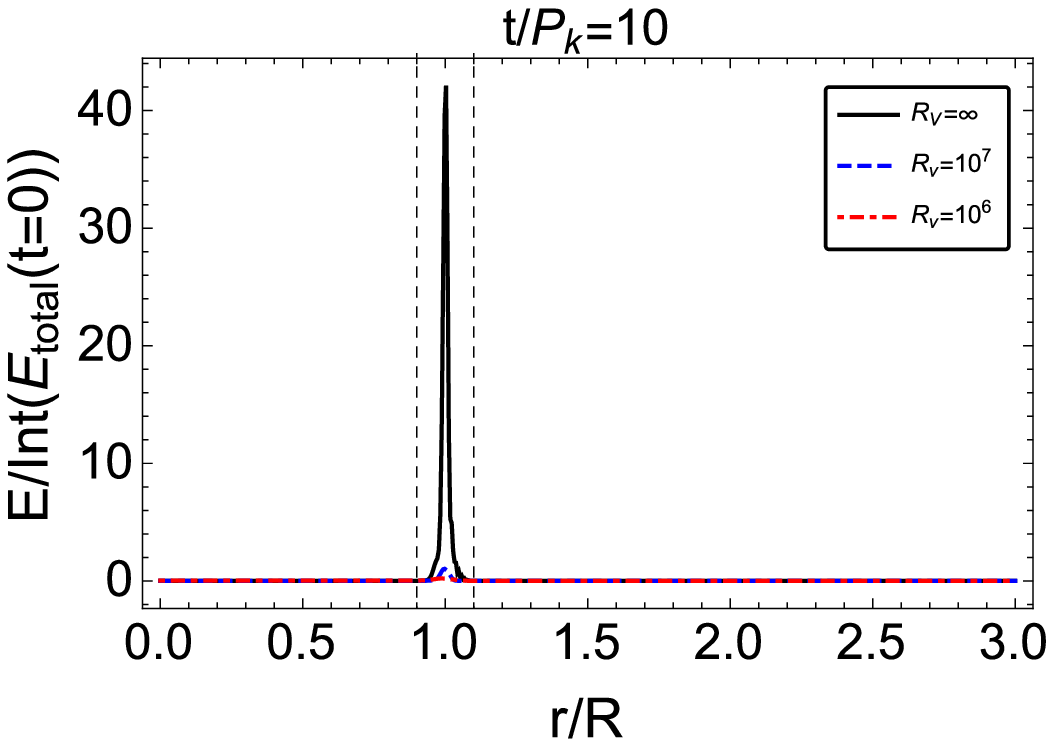}\\
    \includegraphics[width=50mm]{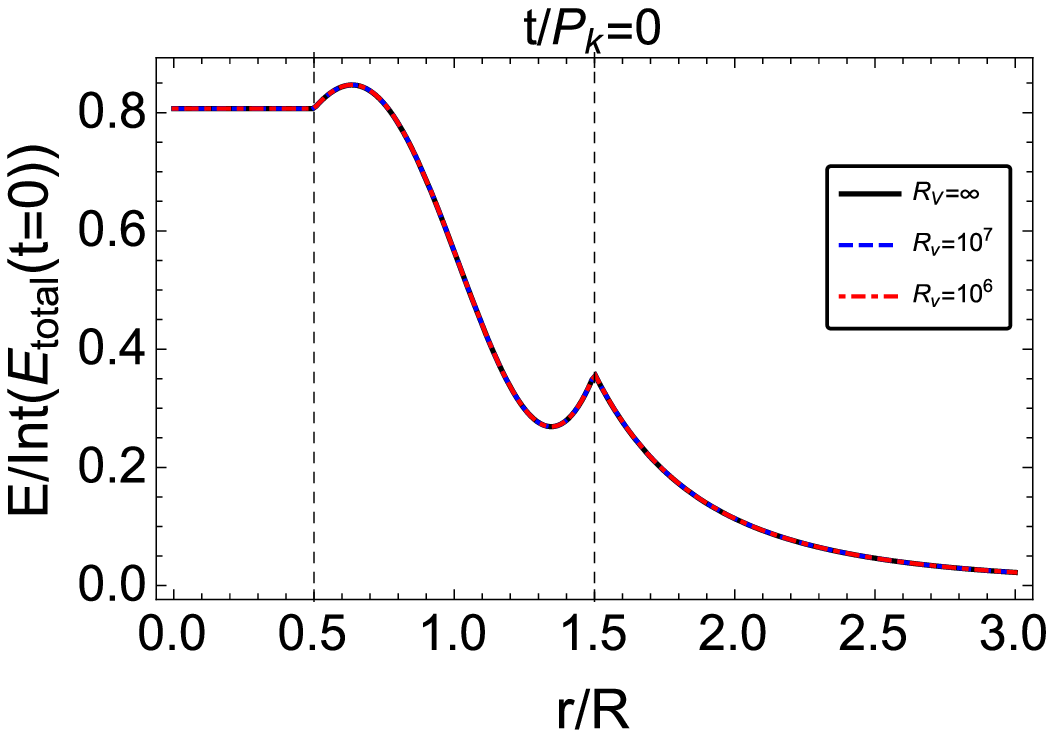} & \includegraphics[width=50mm]{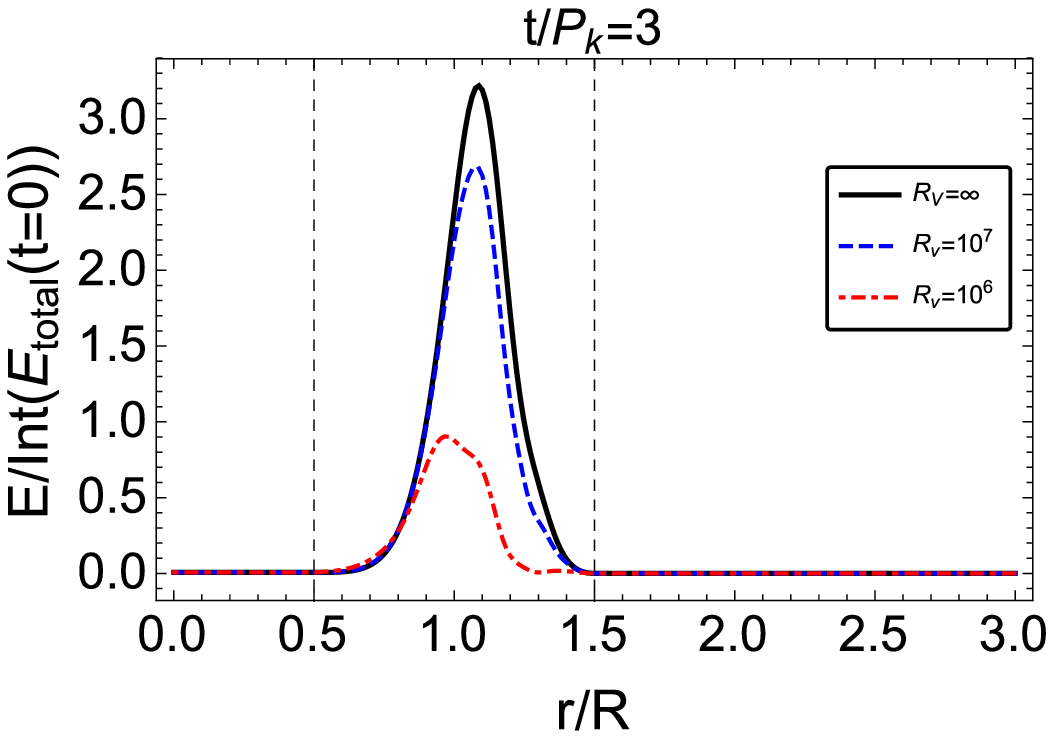}& \includegraphics[width=50mm]{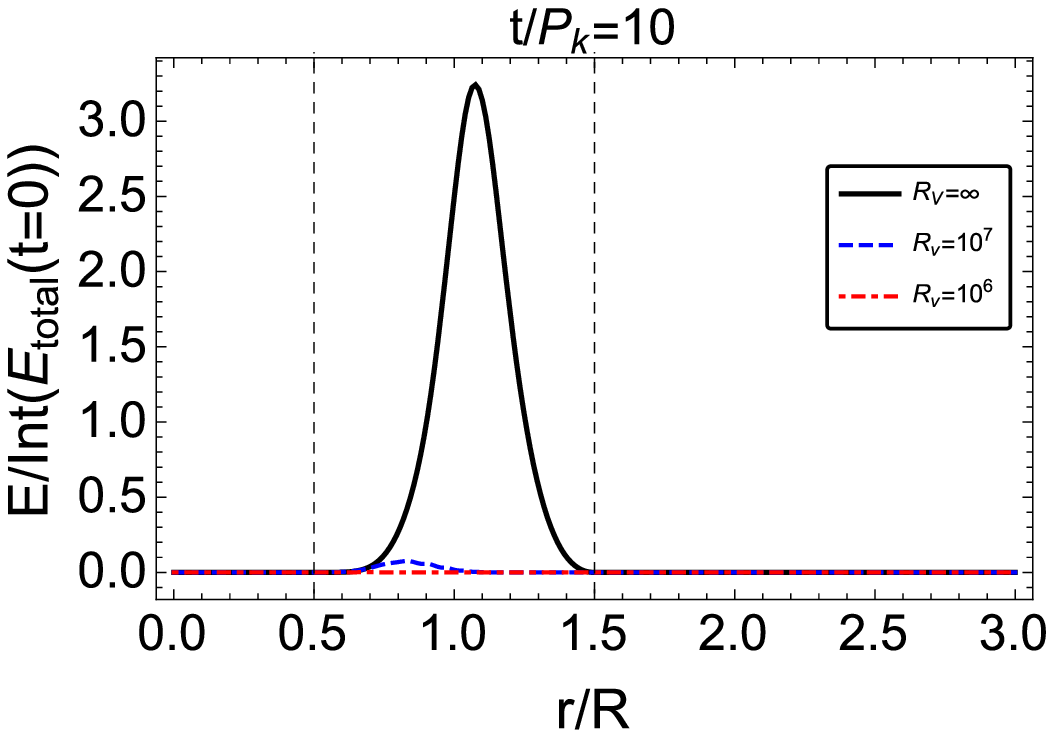} \\
  \end{tabular}
\caption {Energy density as a function of rational radius in unit of the total initial energy with $l/R=0.2$ (top panels) and $l/R=1$ (bottom panels) for $R_v=\infty, 10^6$ and $10^7$. Left, center and right panels denote $t/P_k = 0$, $3$ and $10$, respectively. The left and right vertical dashed lines locate $r=r_1$ and $r=r_2$, respectively. Other auxiliary parameters are as in Figure \ref{spec}.}
    \label{Er}
\end{figure}

\begin{figure}
  \centering
  \begin{tabular}{ccc}
    \includegraphics[width=60mm]{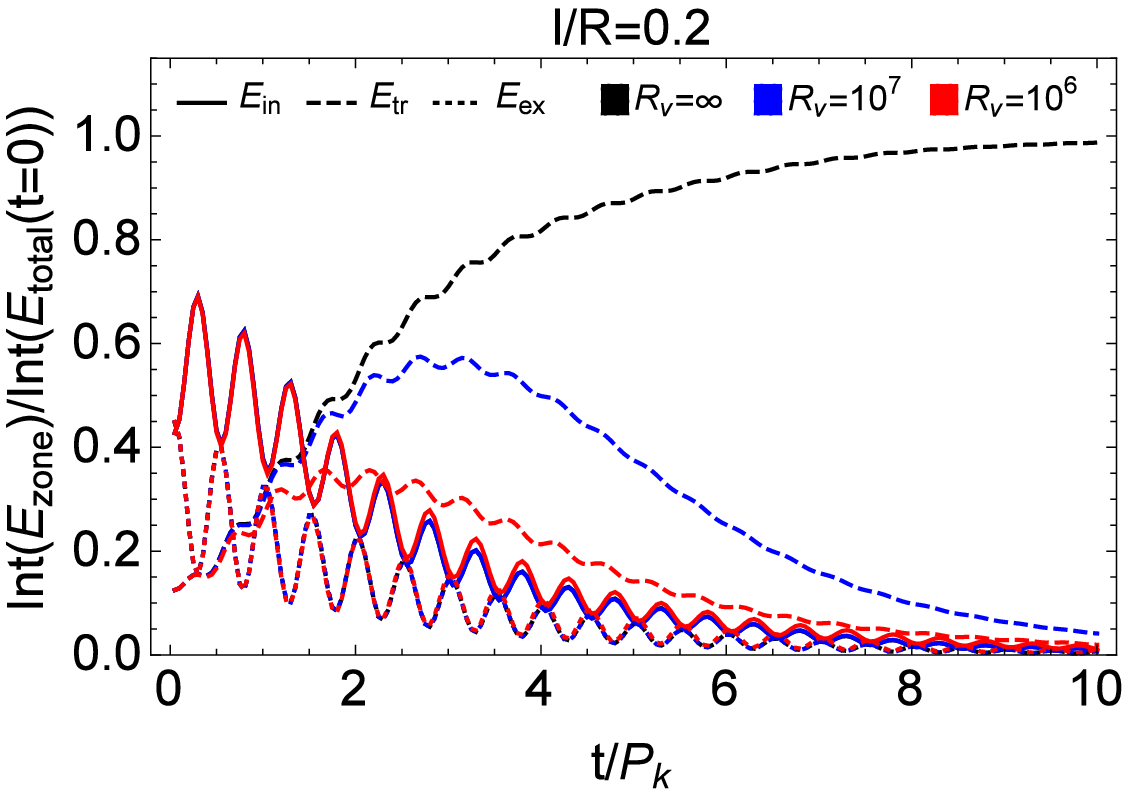} \\
    \includegraphics[width=60mm]{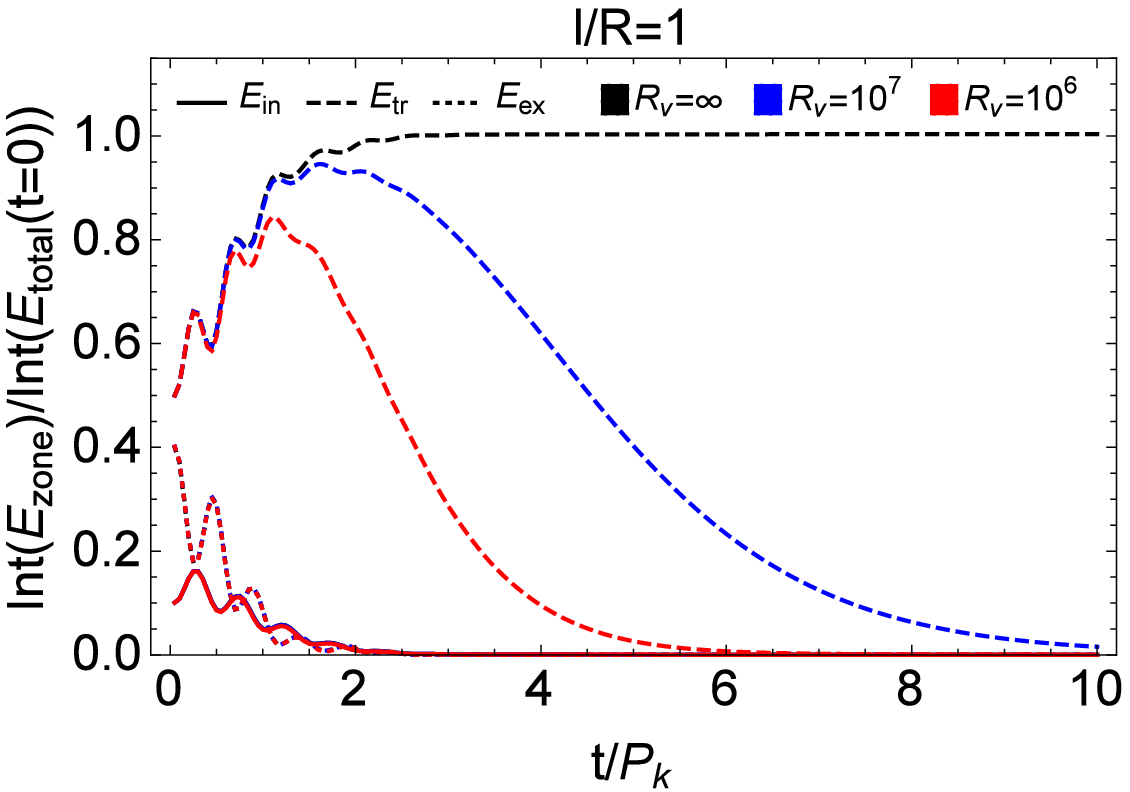}
  \end{tabular}
\caption {Integrated energy of the kink wave in the interior, inhomogeneous region and exterior of the flux tube in unit of the total initial energy versus time for $R_v=\infty, 10^6$ and $10^7$. Top and bottom panel are for $l/R=0.2$ and $l/R=1$, respectively. Other auxiliary parameters are as in Figure \ref{spec}.}
    \label{Ezone}
\end{figure}

\begin{figure}
  \centering
  \begin{tabular}{ccc}
    \includegraphics[width=60mm]{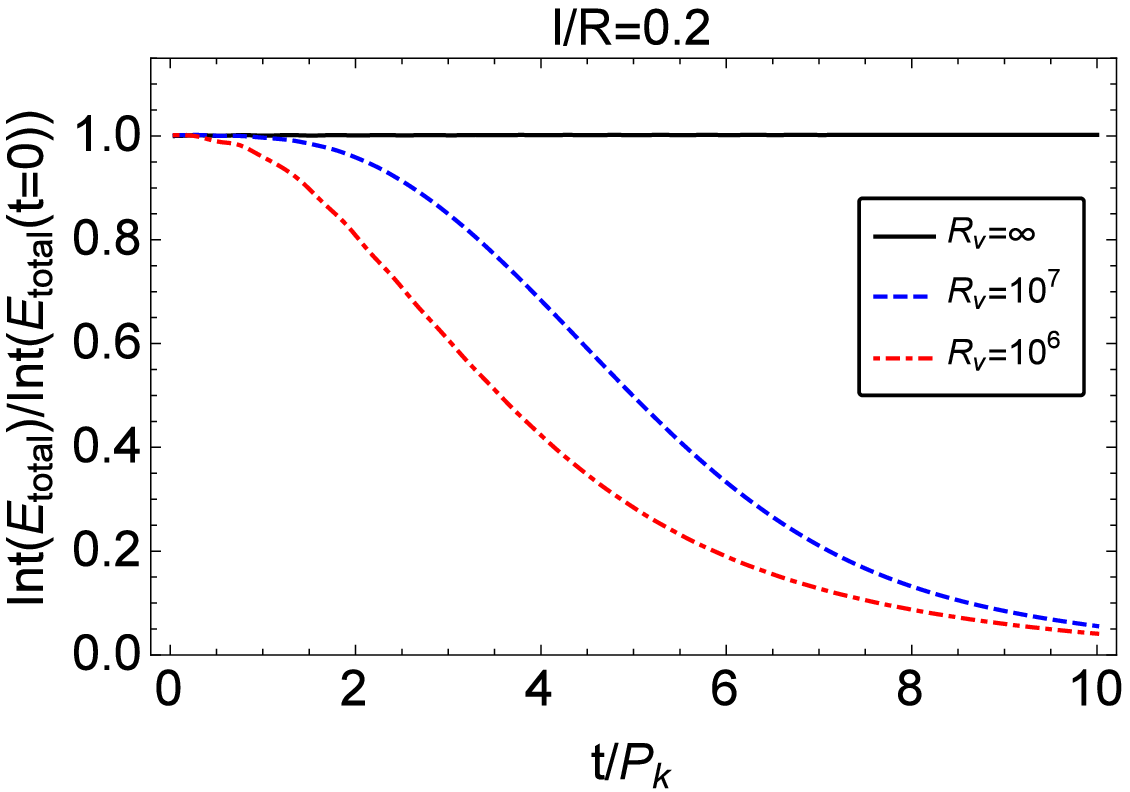} \\
    \includegraphics[width=60mm]{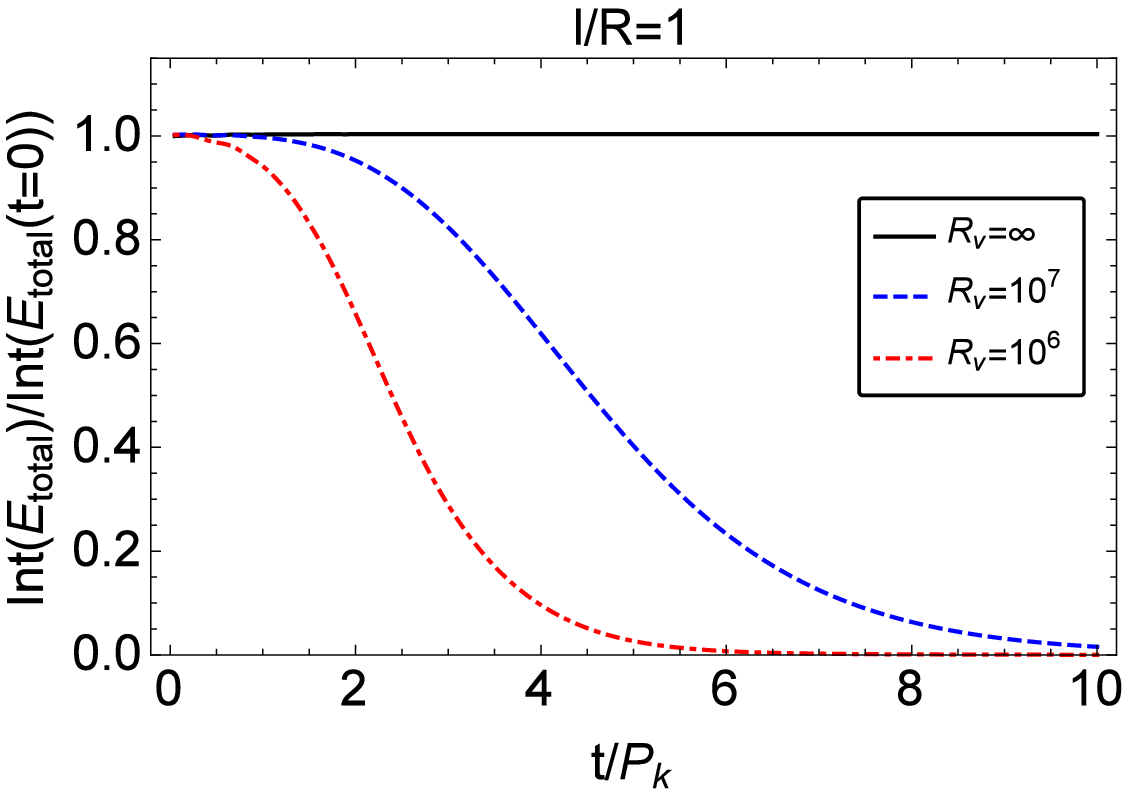}
  \end{tabular}
\caption {Total energy of the kink wave versus time for $R_v=\infty, 10^6 \& 10^7$. Top and bottom panel are for $l/R=0.2$ and $l/R=1$, respectively. Other auxiliary parameters are as in Figure \ref{spec}.}
    \label{Etotal}
\end{figure}

To illustrate the flux of the total energy of the kink wave from the internal and external regions to the inhomogeneous region where it is finally dissipated, we calculate the total energy density of the perturbations as
\begin{equation}\label{med}
     E(r,t)=\frac{1}{2}\left(\rho\left|\frac{\partial \boldsymbol{\xi}}{\partial t}\right|^2+\frac{1}{\mu}\left| \mathbf{B'}\right|^2\right),
\end{equation}
where $\mathbf{B'}$ is obtained from Eq. (\ref{deltaB}).
Figure \ref{Er} illustrates the evolution of the total energy density as a function of $r$. In the absence of viscosity after a few time periods the whole energy of the kink wave concentrates in a narrow layer in the inhomogeneous region of the flux tube. The role of viscosity is to dissipate this concentrated energy. Although, these two processes i.e. flow of energy to the inhomogeneous layer and the dissipation occur simultaneously but they are caused by independent and physically different mechanisms. In order to have a better illustration of the flow of energy of the kink wave from internal and external regions to the inhomogeneous layer and its dissipation, it is appropriate to calculate the integrated total energy of the perturbations in the interior, transitional layer and exterior of the flux tube as a function of time, respectively, as follows
 \begin{eqnarray}\label{Eint}
    \nonumber E_{\rm {in}}=\int_0^{r_1}\frac{1}{2}\left(\rho\left|\frac{\partial \boldsymbol{\xi}}{\partial t}\right|^2+\frac{1}{\mu}\left| \mathbf{B'}\right|^2\right)r~{\rm d}r,\\
    E_{\rm {tr}}=\int_{r_1}^{r_2}\frac{1}{2}\left(\rho\left|\frac{\partial \boldsymbol{\xi}}{\partial t}\right|^2+\frac{1}{\mu}\left| \mathbf{B'}\right|^2\right)r~{\rm d}r,\\
    \nonumber E_{\rm {ex}}=\int_{r_2}^{\infty}\frac{1}{2}\left(\rho\left|\frac{\partial \boldsymbol{\xi}}{\partial t}\right|^2+\frac{1}{\mu}\left| \mathbf{B'}\right|^2\right)r~{\rm d}r.
\end{eqnarray}
Note that in order to compute the third integral of Eq. (\ref{Eint}) we must replace the upper limit of the integral, $\infty$, with a sufficiently large radius where the amplitudes of the perturbations are already negligible. Hence, we check the convergence of the total energy by taking a series of large radii as the upper limit of the third integral in Eq. (\ref{Eint}). Results show that at $r=20R$ the total energy converges to the desired level of accuracy. Figure \ref{Ezone} shows the integrated energy in these three regions as a function of time for $R_v=\infty ,~10^{7} ~\& ~10^{6}$. The solid, dashed and dotted lines denote the integrated energies of the internal, inhomogeneous and external regions, respectively. The black, blue and red colors represent $R_v=\infty$ (ideal MHD), $R_v=10^7$ and $R_v=10^6$, respectively. Note that in the figure, the plots of the internal and external energies (i.e. the solid and dotted lines) for $R_v=\infty$ and $R_v=10^7 \& 10^6$ coincide with each other which reveals that for large Reynolds numbers the existence of viscosity does not affect the energy flow from internal and external regions to the inhomogeneous layer. In other words, the resonant absorption process is not affected by viscosity. Figure \ref{Ezone} also shows that in the presence of viscosity, the energy in the inhomogeneous layer increases in the initial stage of the evolution and reaches a maximum after a few time periods. After that, the energy decreases, since phase mixing in the inhomogeneous layer enhances the viscous dissipation mechanism. The temporal behaviour of the total energy of the kink wave has been illustrated in Figure \ref{Etotal} for $l/R=0.2 ~ \& ~1$ and $R_v=\infty, ~ 10^7 ~\& ~ 10^6$. As figure shows, the energy of the kink wave in the absence of viscosity (black line in the figure) is conserved. The results show that, considering a flux tube with thin transitional layer ($l/R=0.2$), for $R_v=10^7~\&~10^6$ the energy of the kink wave decreases to $1/e$ of its initial energy after $t\simeq 5.8P_k$ and $t\simeq 4.4P_k$, respectively. For thick transitional layer ($l/R=1$) the corresponding values are $t\simeq 5.2P_k$ and $t\simeq 2.7P_k$.

\begin{figure}
  \centering
  \begin{tabular}{ccc}
    \includegraphics[width=60mm]{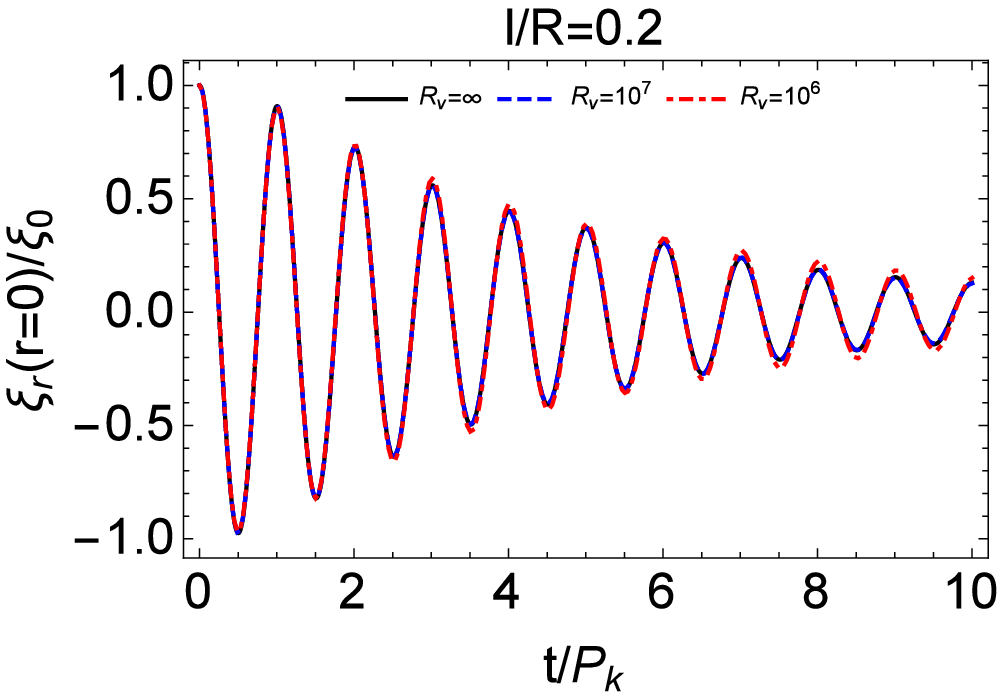} \\
    \includegraphics[width=60mm]{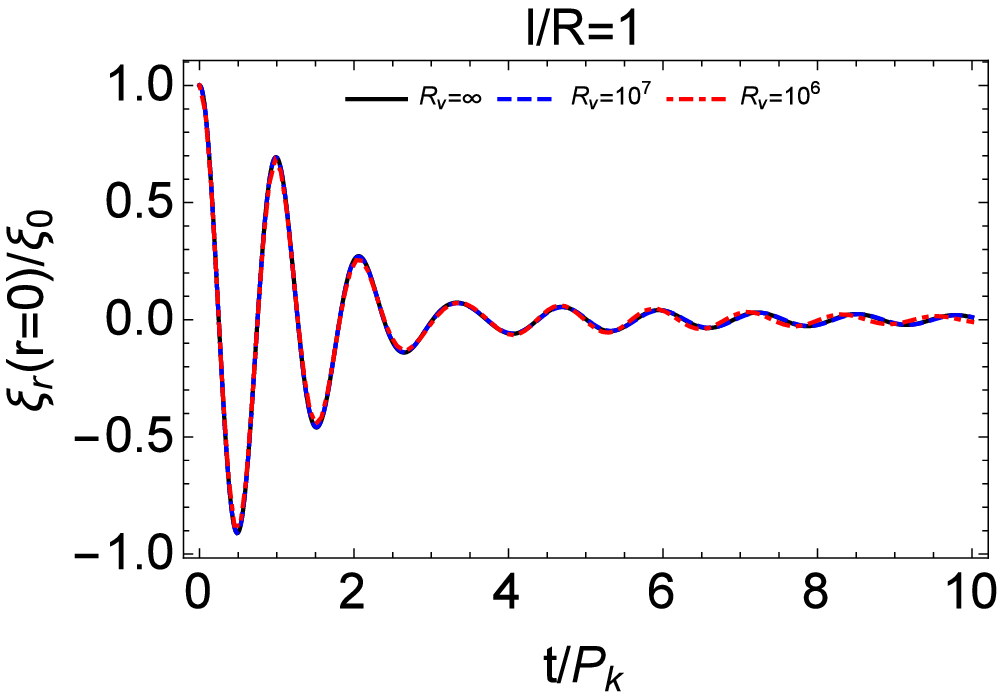}
  \end{tabular}
\caption {Temporal behavior of $\xi_r$ on the axis of the flux tube for $R_v=\infty, 10^6$ and $10^7$. Top and bottom panel are for $l/R=0.2$ and $l/R=1$, respectively. Other auxiliary parameters are as in Figure \ref{spec}.}
    \label{xirt}
\end{figure}

Figure \ref{xirt} shows the radial component of the displacement at the axis of the flux tube. Interestingly the results for $R_v=\infty, 10^7 \& 10^6$ are the same. Note that in the observations of kink waves in coronal flux tubes, we measure the displacement of the axis but detecting the rotational motions related to the kink waves in the inhomogeneous boundary of the flux tubes is not an easy task. Comparing the results illustrated in Figures \ref{Etotal} and \ref{xirt}, it is clear that although the temporal behaviour of the displacement of the flux tube axis is the same in both cases of ideal and viscous MHD (with large Reynolds number), the total energy of the kink waves which is conserved for $R_v=\infty$, decays in a few periods for $R_v=10^7$ and $10^6$. Hence, one can conclude that assuming a large Reynolds number for the coronal plasma the global damping of kink waves reported in the observations is due to converting the transverse motion to rotational perturbations in the inhomogeneous layer of the flux tube, i.e. the resonant absorption process, and is not related to the existence of viscosity.

\begin{figure}
  \centering
  \begin{tabular}{ccc}
    \includegraphics[width=70mm]{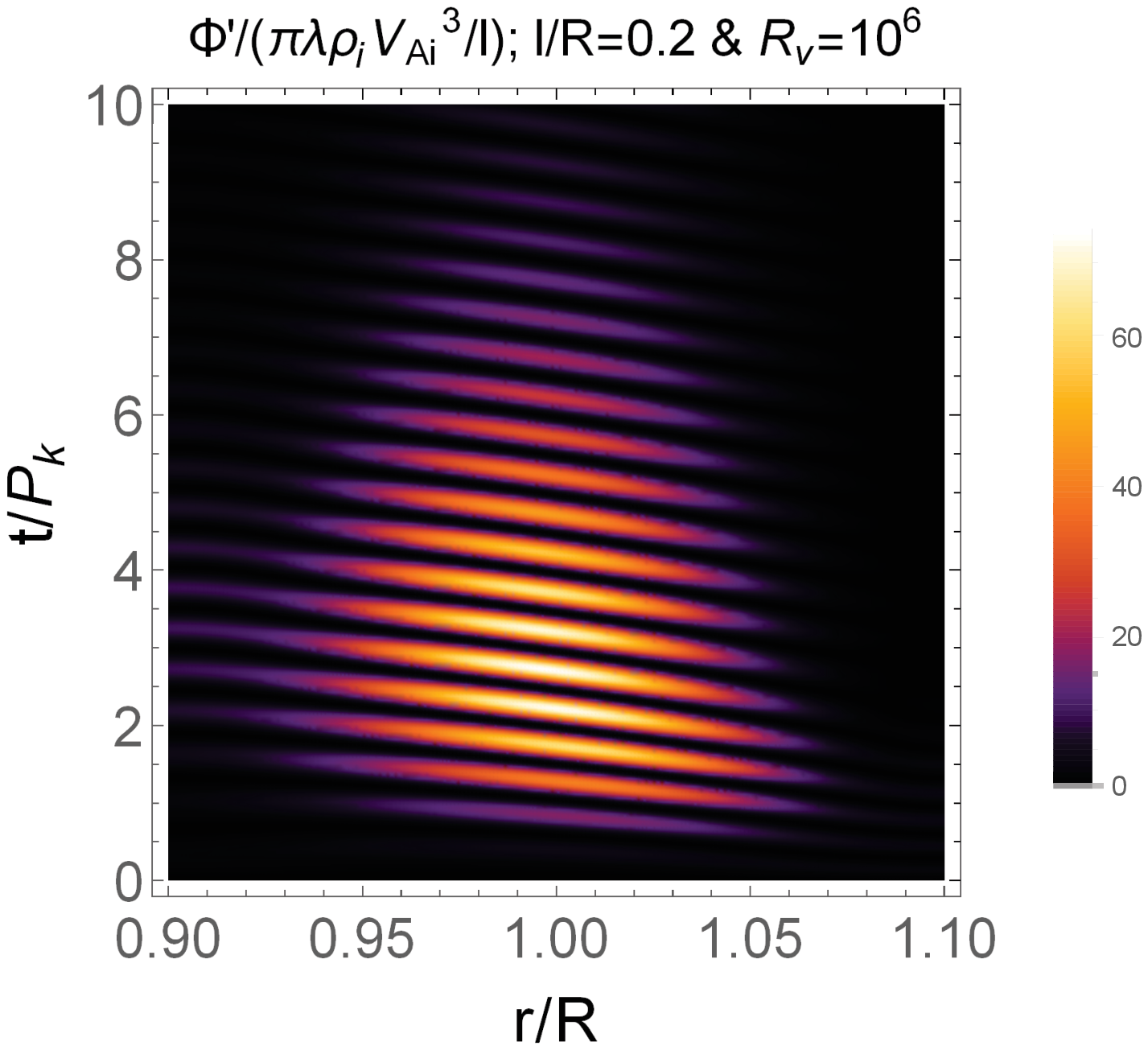}& \includegraphics[width=66mm]{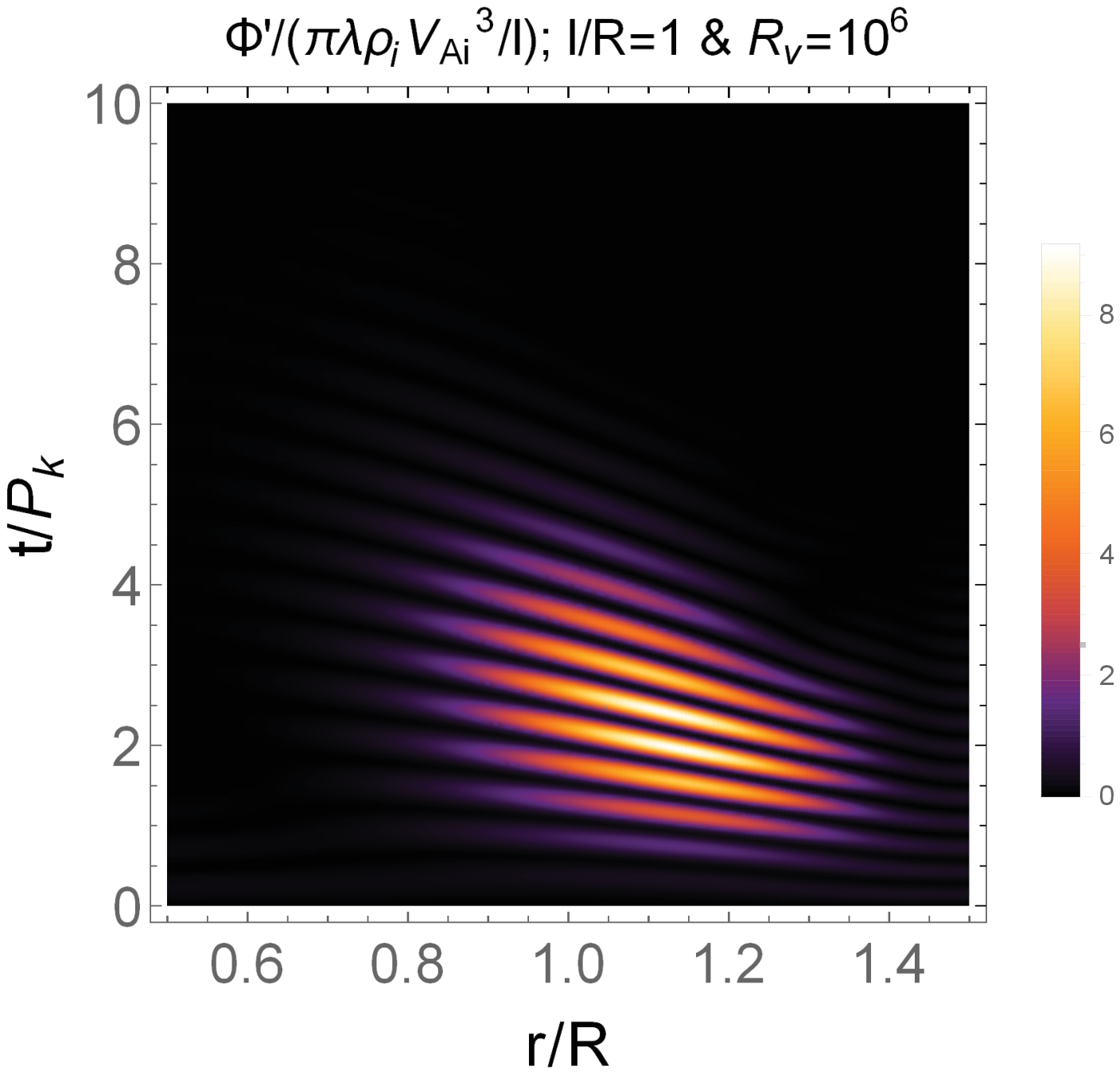} \\
    \includegraphics[width=70mm]{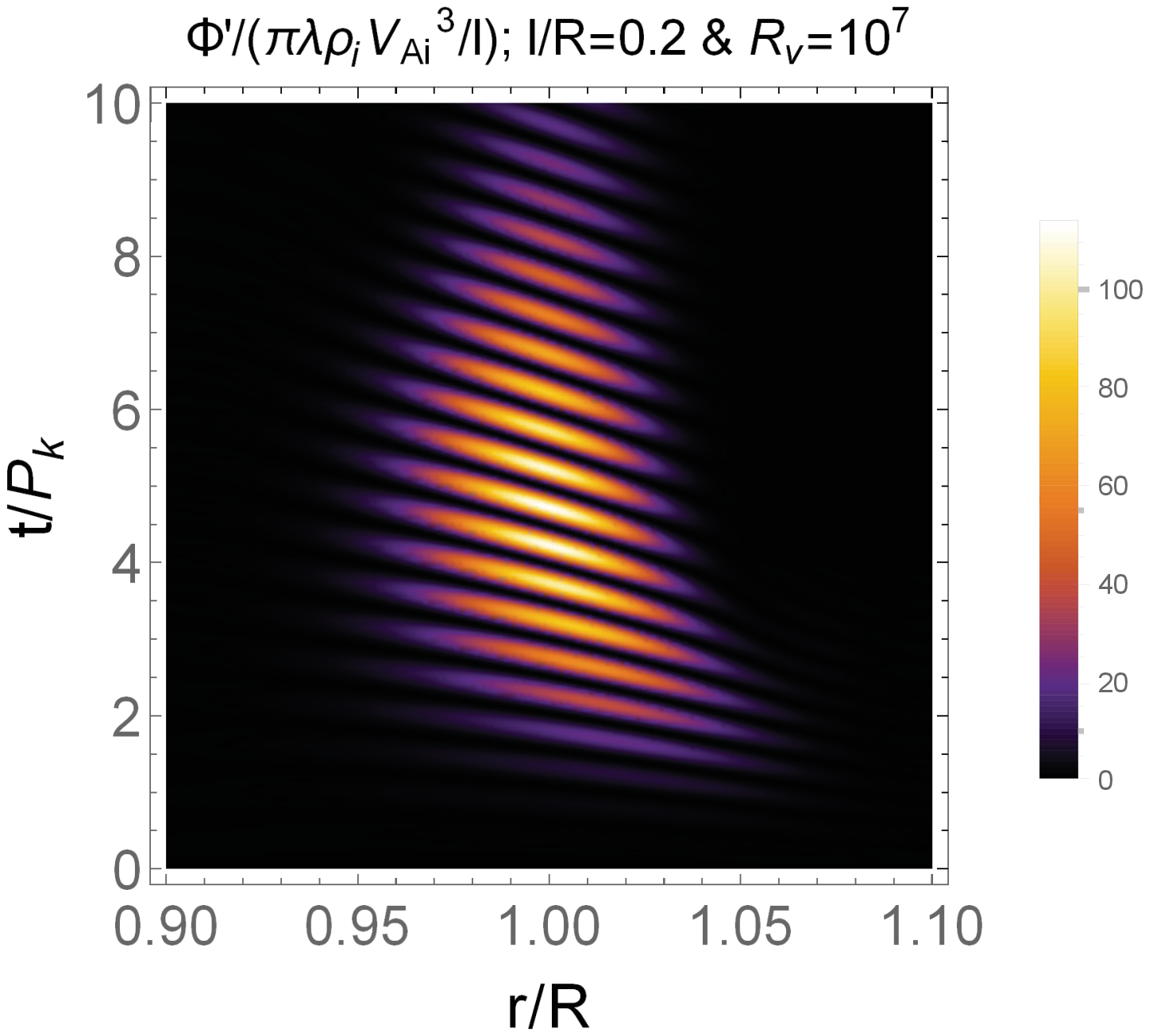}& \includegraphics[width=66mm]{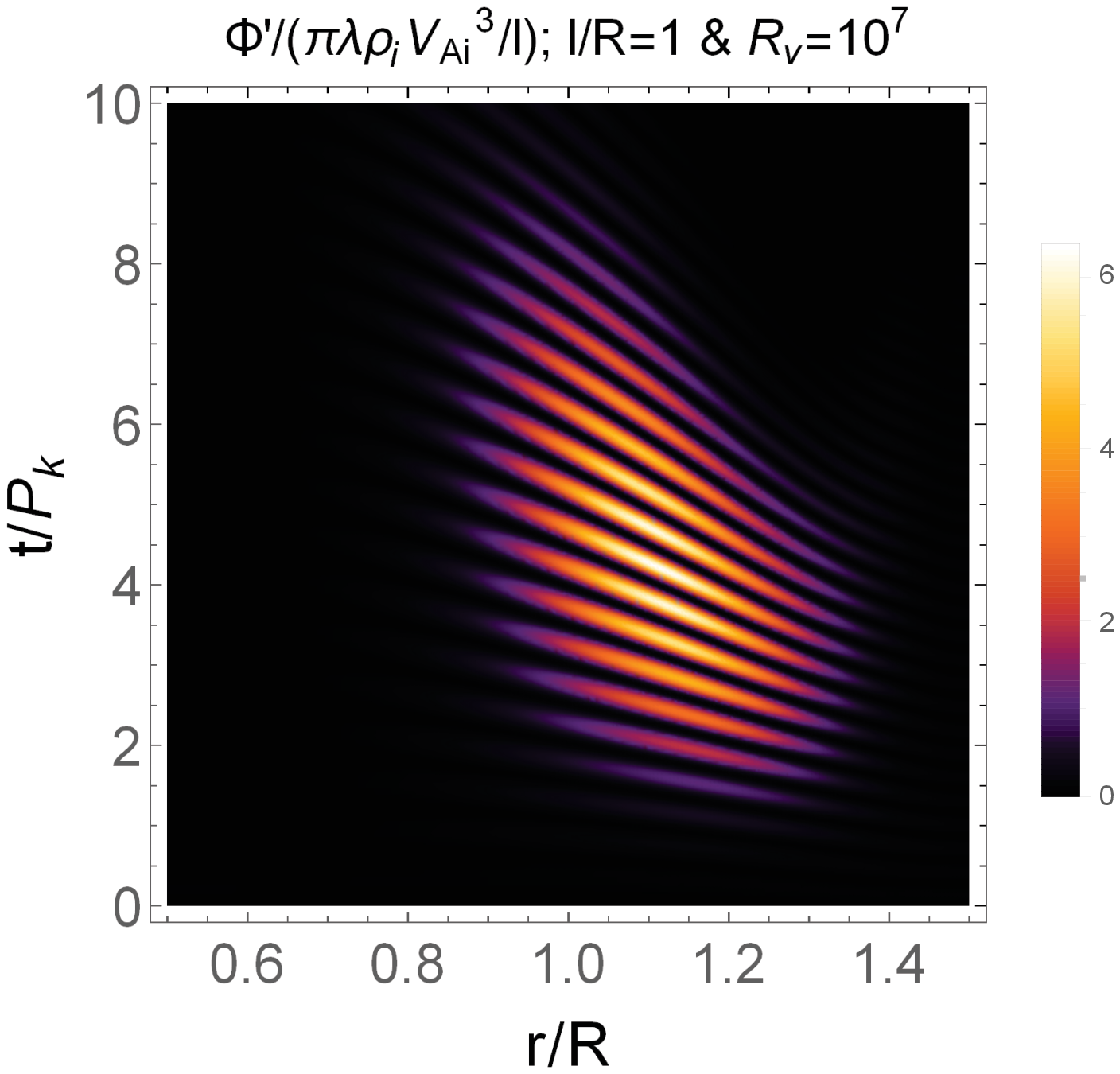} \\
  \end{tabular}
\caption {Contours of the dissipation function in normalized units in $r-t$ plane for $R_v= 10^6$ (top panels) and $10^7$ (bottom panels). Left and right panels are for $l/R=0.2$ and $l/R=1$, respectively. Other auxiliary parameters are as in Figure \ref{spec}. }
    \label{HR}
\end{figure}

In order to illustrate the rate of plasma heating in the inhomogeneous region due to viscosity we use the viscous dissipation function that for an incompressible plasma with constant viscosity is as follows (White 1991)
\begin{equation}\label{Phi}
  \Phi=\rho\nu\left[2\left(\varepsilon_{rr}^2+\varepsilon_{\varphi\varphi}^2+\varepsilon_{zz}^2\right)+\varepsilon_{r\varphi}^2+\varepsilon_{rz}^2+\varepsilon_{\varphi z}^2\right],
\end{equation}
where $\varepsilon_{ij}$ with $i, j=r, \varphi, z$ are components of the strain rate tensor defined as follows

\begin{eqnarray}\label{epsilons}
  \nonumber\varepsilon_{rr}&=&\frac{\partial^2\xi_r}{\partial r \partial t},\\
  \nonumber\varepsilon_{\varphi\varphi}&=&\frac{1}{r}\frac{\partial^2\xi_\varphi}{\partial \varphi \partial t}+\frac{1}{r}\frac{\partial \xi_r}{\partial t},\\
  \nonumber\varepsilon_{zz}&=&\frac{\partial^2\xi_z}{\partial z \partial t},\\
  \nonumber\varepsilon_{r \varphi}&=&\frac{1}{r}\frac{\partial^2\xi_r}{\partial \varphi \partial t}+\frac{\partial^2 \xi_\varphi}{\partial r \partial t}-\frac{1}{r}\frac{\partial \xi_\varphi}{\partial t},\\
  \nonumber\varepsilon_{r z}&=&\frac{\partial^2\xi_r}{\partial z \partial t}+\frac{\partial^2 \xi_z}{\partial r \partial t},\\
  \nonumber\varepsilon_{\varphi z}&=&\frac{1}{r}\frac{\partial^2\xi_z}{\partial \varphi \partial t}+\frac{\partial^2 \xi_\varphi}{\partial z \partial t}.
\end{eqnarray}
The dissipation function, $\Phi$, is the work done by the viscous stresses on an element of plasma per unit volume per unit time. Here, we remove the $\varphi$ and $z$ dependency of the dissipation function by integrating $\Phi$ in these directions, i.e.
\begin{equation}\label{Phiprime}
  \Phi'=\int_{0}^{\lambda}\int_{0}^{2\pi}\Phi r d\varphi dz=\pi\lambda r\Phi,
\end{equation}
where, $\lambda=2\pi/k_z$ is the wavelength. Figure \ref{HR} shows the contour plot of $\Phi'$ in $r-t$ plane for $l/R=0.2$ and $l/R=1$ with $R_v=10^6$ and $R_v=10^7$. As figure shows, the heating rate has an oblique oscillatory pattern. The obliquity of the contours of $\Phi'$ is due to the phase mixing of the perturbations. In order to obtain the time that the dissipation in the inhomogeneous region reaches its maximum we have calculated the integral of $\Phi'$ over the range $[r_1, r_2]$ that is a function of time. For $R_v=10^6$, considering $l/R=0.2$ and $l/R=1$ we obtain the peak time of the integrated dissipation as $t_{peak}=2.75 P_k$ and $t_{peak}=2.1 P_k$, respectively. The corresponding values for $R_v=10^7$ are $t_{peak}=4.5P_k$ and $t_{peak}=4.1P_k$. For $l/R=0.2$ and $R_v=10^8$ we get $t_{peak}=8P_k$. So, for the larger value of the Reynolds number the system needs to be more phase mixed for viscosity to reach its maximum efficiency as a heating mechanism for the plasma. Obviously, if the Reynolds number is further increased to the expected value in the solar corona,
the peak efficiency of dissipation happens in a much later time corresponding to many periods of the original kink oscillation.
Since the observationally reported damping time of kink oscillations corresponds to a few periods (as the theory of resonant absorption
correctly predicts), we conclude that viscous heating becomes of significance only after the global kink oscillation is damped, i.e., no heating is expected during the damping of the global kink oscillation in realistic coronal conditions.
%________________________________________________________________________________________________________

\section{Summary and Conclusions}\label{Conclusions}

Here, we investigated the effect of viscosity on the evolution of MHD kink waves in coronal flux tubes.
We modelled a magnetic flux tube by a straight magnetic cylinder. Plasma density inside and outside the tube is constant with different values. The interior and exterior of the tube are connected by an inhomogeneous transitional layer in which the plasma density varies smoothly with a sinusoidal profile from the internal value to the external one. The background magnetic field is aligned with tube axis and has constant
magnitude everywhere. We neglected the role of viscosity in the constant density regions since in the limit of small viscosities the dissipation is only important in the inhomogeneous region where the phase mixing process is at work. Using the modal expansion technique (Cally 1991; Soler \& Terradas 2015) we solved the viscous MHD equations of motion in thin tube approximation and obtained the spatio-temporal behaviour of the perturbations in the flux tube. We considered both the cases of thin and thick inhomogeneous layers in our analysis.

We obtained the spectrum of the complex eigenfrequencies of the Alfv\'{e}n discrete modes in the inhomogeneous layer. In the spectrum, one of the eigenfrequencies could be identified as the quasi-mode solution of the kink waves by the resonant absorption mechanism.
\newpage
To investigate the effect of viscosity on the kink waves, we obtained the temporal and spatial behaviour of the perturbations in the interior, inhomogeneous region and exterior of the flux tube. Our results showed that for both cases of thin and thick inhomogeneous layers, considering the viscosity in the system does not affect the transverse motion of the flux tube axis and its decay rate for large Reynolds numbers ($R_v\geq 10^6$) confirming the previously results obtained by e.g. Ruderman \& Roberts (2002) and Goossens et al. (2002).  This result confirms that the fast damping of the kink oscillations in coronal loops could be a consequence of resonant absorption mechanism which despite the existence of viscosity naturally results to changing the behaviour of kink mode from mainly  transverse motion to rotational motion (Goossens et al. 2014). However, even for small viscosities (relevant for the coronal plasma) the viscous dissipation is important in the developed stage of phase mixing of the perturbations in the inhomogeneous region. Our results show that viscosity eventually suppresses the rate of phase mixing of the perturbations by coupling the neighboring magnetic surfaces and transforming their energy to heat.

In order to investigate the effect of viscosity on cascading the total (kinetic plus magnetic) energy of the kink wave to the inhomogeneous layer of the flux tube, we obtained the total energy density as well as the integrated total energy as a function of $r$ and $t$. As in the ideal MHD case, in presence of viscosity the energy of the kink wave tends to concentrate in a narrow region in the inhomogeneous layer of the flux tube but it is not allowed to achieve its maximum value obtained in the ideal case since the dissipation mechanism is at work and decreases the energy in time. However, the small amount of viscosity considered in our analysis does not affect the energy flow from interior and exterior of the tube to the inhomogeneous region. Temporal behavior of the total energy showed that for both the thin and thick inhomogeneous region with Reynolds numbers of the order $10^6$, $10^7$ and $10^8$ the energy of the kink wave decays to heat the plasma within a few periods. However, if larger and more realistic values of the Reynolds numbers were used, heating would happen much after the observable kink oscillation is completely damped. These conclusions agree with the simple estimations provided in the research note by Terradas \& Arregui (2018), although these authors considered resistive heating instead of viscous heating.

We also studied the efficiency of heating duo to viscosity by calculating the dissipation function in the inhomogeneous region. The obtained results show that at initial stage of the evolution the dissipation increases with time and reaches a maximum level after a few periods (2-8 periods for $R_v=10^6-10^8$ but much later for realistic $R_v$) in a narrow layer near the boundary of the flux tube. After that the dissipation decreases since the energy budget provided by the initial value problem considered in this paper is finite.

In summary, we showed that viscosity, even in a small amount, can have a significant impact on the later evolution of phase mixing by the suppressing the generation of small scales and transforming the energy of the wave to heat. Reynolds number larger than the values considered in this paper needs more massive numerical computation with the mathematical approach presented here to obtain the correct set of complex eigenvalues of the damped Alfv\'{e}n discrete modes which could be a subject of future work.

We finally note that this work is based on linear theory. Nonlinear effects may modify somehow these results, since presumably important ingredients as, e.g., the triggering of Kelvin-Helmholtz instabilities are absent from our study (see e.g. Terradas et al. 2008). The nonlinear evolution should be necessarily investigated with high-resolution dissipative MHD simulations.

%________________________________________________________________________________________________________
\section*{Acknowledgements}
RS acknowledges the support from grant AYA2017- 85465-P (MINECO/AEI/FEDER, UE) and from the Ministerio de Economía, Industria y Competitividad and the Conselleria d’Innovació, Recerca i Turisme del Govern Balear (Pla de ciència, tecnologia, innovació i emprenedoria 2013−2017) for the Ramón y Cajal grant RYC-2014-14970.
%________________________________________________________________________________________________________

%-----------------------------------------------------------------------------------------------
%\clearpage
%\begin{figure}
%\special{psfile=Bfi.eps vscale=100 hscale=100 hoffset=0 voffset=0 }
% \vspace{0cm}
%\caption[] {The azimuthal component of magnetic field. The maximum
%twist occurs at the mid point of the surface layer. The blue curve
%is the parabolic function of equation (\ref{Bphi}) and the red
%vectors represent $B_\varphi$ qualitatively. } \label{Bfi}
% \end{figure}
%-----------------------------------------------------------------------------------------------
%\clearpage
%\begin{figure}
%\special{psfile=fig1.eps vscale=100 hscale=100 hoffset=0 voffset=0 }
% \vspace{0cm}
%\caption[] {Background Alfv\'{e}n Frequency for $m=1$, $l=1$,
%$\frac{a}{R}=0.99$, $\frac{R}{L}=0.01$, $\frac{\rho_{\rm
%i}}{\rho_{\rm e}}=2$, $\rho_{\rm i}=2\times10^{-14} \frac{kg}{m^3}$,
%$B_{\rm i}=100 {\rm G}$ and $\alpha=\alpha_j=0.013$(solid),
%$0.015$(dashed), $0.011$(dash-dotted).} \label{omegaA}
% \end{figure}
%-----------------------------------------------------------------------------------------------

%-----------------------------------------------------------------------------------------------

%-
\end{document}